\documentclass{ieeeaccess}
\usepackage{cite}
\usepackage{amsmath,amssymb,amsfonts}
\usepackage{algorithmic}
\usepackage{graphicx}
\usepackage{textcomp}

\usepackage{enumerate}
\usepackage{subcaption}
\usepackage{algorithm}
\usepackage{stfloats}
\usepackage{caption}
\usepackage{multirow}
\usepackage{color}
\usepackage{gensymb}
\usepackage{amsfonts}
\usepackage{soul,color}
\usepackage{hyperref}
\usepackage{soul}
\def\BibTeX{{\rm B\kern-.05em{\sc i\kern-.025em b}\kern-.08em
    T\kern-.1667em\lower.7ex\hbox{E}\kern-.125emX}}
\begin{document}
\history{Date of publication xxxx 00, 0000, date of current version xxxx 00, 0000.}
\doi{10.1109/ACCESS.2019.2923174}

\title{PreLatPUF: Exploiting DRAM Latency Variations for Generating Robust Device Signatures}

\author{\uppercase{B. M. S. Bahar Talukder}\authorrefmark{1}, \IEEEmembership{Student Member, IEEE},
\uppercase{Biswajit Ray}\authorrefmark{1}, \IEEEmembership{Member, IEEE},
\uppercase{Domenic Forte}\authorrefmark{2}, \IEEEmembership{Senior Member, IEEE}, and
\uppercase{Md Tauhidur Rahman}\authorrefmark{1}, \IEEEmembership{Member, IEEE}}

\address[1]{ECE Department, University of Alabama in Huntsville, Huntsville, AL 35899 USA (e-mail: \{bms.btalukder, biswajit.ray, tauhidur.rahman\}@uah.edu)}
\address[2]{ECE Department, University of Florida, Gainesville, FL 35899 USA (e-mail: dforte@ece.ufl.edu)}

\tfootnote{This work was supported in part by the National Science Foundation under Grant Number CNS-1850241 and UAH.}

\markboth
{B. M. S. B. Talukder \headeretal: PreLatPUF: Exploiting DRAM Latency Variations for Generating Robust Device Signatures}
{B. M. S. B. Talukder \headeretal: PreLatPUF: Exploiting DRAM Latency Variations for Generating Robust Device Signatures}

\corresp{Corresponding author: B. M. S. Bahar Talukder (e-mail: bms.btalukder@uah.edu).}

\begin{abstract}
Physically Unclonable Functions (PUFs) are potential security blocks to generate unique and more secure keys in low-cost cryptographic applications. Dynamic random-access memory (DRAM) has been proposed as one of the promising candidates for generating robust keys. Unfortunately, the existing techniques of generating device signatures from DRAM is very slow, destructive (destroy the current data), and disruptive to system operation. In this paper, we propose \textit{precharge} latency-based PUF (PreLatPUF) that exploits DRAM \textit{precharge} latency variations to generate signatures. The proposed PreLatPUF is fast, robust, least disruptive, and non-destructive. The silicon results from commercially available $DDR3$ chips from different manufacturers show that the proposed key generation technique is at least $ \sim 1,192X$ faster than the existing approaches, while reliably reproducing the key in extreme operating conditions.
\end{abstract}


\titlepgskip=-15pt

\maketitle

\section{Introduction}
\label{sec:introduction}
\PARstart{P}{hysical} unclonable functions (PUFs) play important roles in security by offering a high level of protection in cryptographic applications with the capability of strong volatile key or unique ID generation. A PUF is a circuit that generates unique fingerprints by exploiting the inherent and unavoidable manufacturing process variations during fabrication \cite{Koushanfar:PUFapp, Keller:DRAMPUF}. Identification, authentication, secure communication, IC obfuscation to prevent IC piracy in semiconductor supply chain, detection of counterfeit ICs, etc. are a few common applications of PUFs because of their unique and unpredictable characteristics \cite{guin:IPsec,Chakraborty:HARPOON, SRAM-PUF:Tauhid, Koushanfar:PUFapp, CSST, SST, pira:bhunia, jtag:bhunia}. In recent years, PUFs have also been used in IoT applications because they enable low-cost solutions with a high level of security \cite{PUF_IOT_1,PUF_IOT_2,PUF_IOT_3}. 

In addition to low-cost, the memory-based PUF provides an opportunity to implement PUF-based schemes to the existing system \cite{SRAM-PUF:Tauhid, Keller:DRAMPUF, pira:bhunia, jtag:bhunia}. The start-up behavior of the memory chips, disturbance characteristics, the random decay properties, etc. are the most common techniques to generate responses from memory chips \cite{Koushanfar:PUFapp}. Previous works on DRAM PUFs (DPUFs) have focused on: (i) retention-based: writing all cells to `1' and disabling the refresh then waiting for half the cells to discharge and reading cell values \cite{Keller:DRAMPUF,retentionPUF, retentionPUF2, retentionPUF3}, (ii) start-up based: using the start-up values of the cells to generate the secret key as in \cite{Sarah:DRAMPUF,DPUFstartup}, and (iii) disturbance-based: disturbance caused by rowhammer \cite{Schaller:hammerPUF, Anagnostopoulos:hammerPUF}. The variations in \textit{activation} latency time have also been used to generate device signatures \cite{DRAMLatencyPUF}. In this method, the signature is obtained from the errors generated at the reduced \textit{activation} time during read operation \cite{DRAMLatencyPUF}. 

In PUF-based applications, the responses (i.e., the PUF outputs) have to be robust, fast, random, and unique \cite{SRAM-PUF:Tauhid, MemristorPUF, KanSRAMPUF, ARO-PUF:Tauhid,ARO-PUFJournal}. Like other silicon PUFs, the DRAM-based PUF responses are also impacted by external influences such as operating and environmental variations, aging, etc. \cite{reducedvoltageOP, processMargine, designInducedTimingVar, aging, hci, Ganta2014, mosfetTempEffect, scare:guo}. In addition, the existing signature generation schemes from DRAM do not offer impressive throughput; retention-based DPUF requires an order of minutes, and start-up based DPUF needs a power cycle. The destructiveness of the memory contents, disruption of the system, etc. are few other major limitations of existing DRAM-based PUFs (discussed in Section \ref{Motivations}).

While some applications can tolerate a certain amount of errors, others, such as the generation of cryptographic keys, cannot. To make the PUF output more stable (i.e., to obtain the same response for the applied challenge to a PUF), error correcting code (ECC) and different enrollment schemes are often used but at the expense of additional cost \cite{CIBS-ECC, ECCPUF, ECC_compare}.

In this paper, we propose PreLatPUF that exploits the \textit{precharge} timing latency variations in DRAM to generate device signatures. The main contributions of this paper (i.e., to generate robust device signatures from DRAM) are summarized below.

\begin{itemize}
\item We propose \textit{precharge} latency based DRAM PUF (PreLatPUF) that generates device signatures at a much faster rate. We experimentally demonstrate that the faulty read operation at the reduced \textit{precharge} latency can be used to generate unique and random device signatures.
\item We characterize the errors at the reduced \textit{precharge} latency to discover cells that are most suitable for robust and reliable PUFs. 
\item We propose a cell selection algorithm and a registration technique to ensure that the signatures generated at the reduced \textit{precharge} latency are robust, unique, and random. 
\item We present a quantitative and qualitative comparison between PreLatPUF and some of the previously proposed DRAM-based PUFs. The results show that the proposed PreLatPUF outperforms existing DPUFs in several aspects. 
\item We evaluate the proposed PreLatPUF using commercially available DDR3 DRAM modules. 
\end{itemize}

The rest of the paper is organized as follows. In Section \ref{sec:background}, we present the background of DRAM architecture, read/write operation, existing DRAM-based PUFs and major challenges. We propose the latency-based DRAM PUF in Section \ref{sec:designoverview}. The experimental results and discussions are presented in Section \ref{sec:results}. We conclude the paper in Section \ref{sec:conclusion}.

\section{Background and Motivation} \label{sec:background}
In this section, we provide a brief background of the modern memory subsystem and its operation. We also present existing DRAM-based PUFs and their limitations.
 
\subsection{DRAM Organization}

\begin{figure*}[ht!]
\centering
\begin{subfigure}[t!]{0.59\textwidth}
\includegraphics[width=\textwidth]{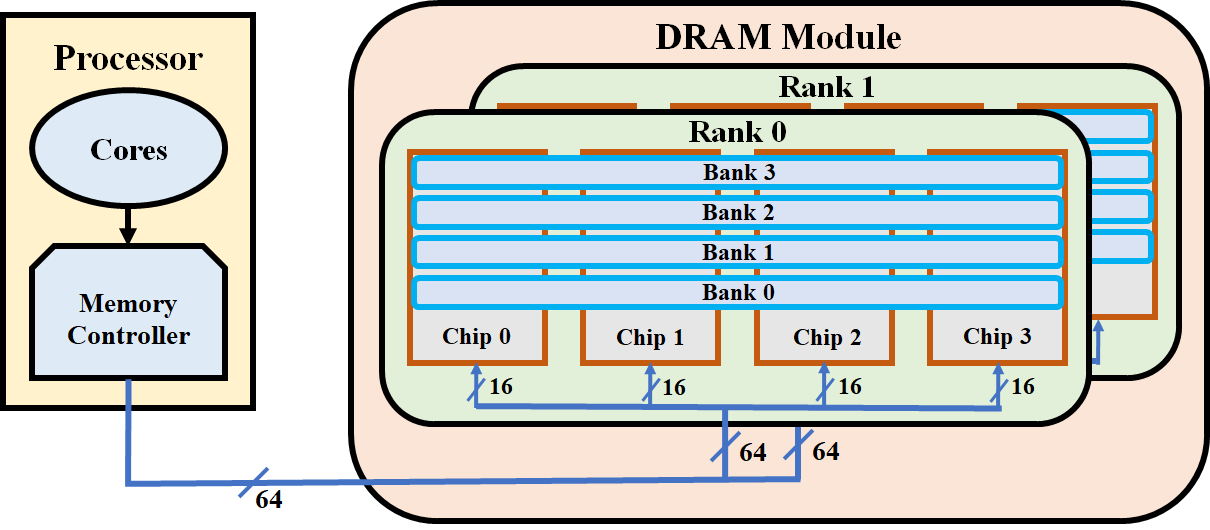}
\caption{DRAM system.}
\label{fig:blockDiagram}
\end{subfigure}
~
\begin{subfigure}[t!]{0.39\textwidth}
\includegraphics[width=\textwidth]{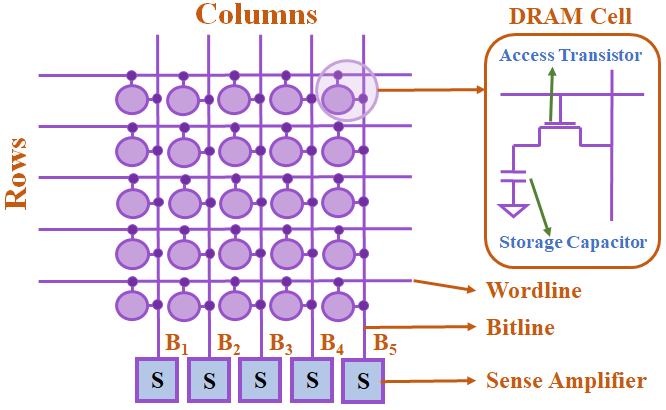}
\caption{Close view on a Memory Bank.}
\label{fig:DRAMcell}
\end{subfigure}
\captionsetup{font={sf,small},labelfont={bf,color=accessblue}}
\caption{Organization of a modern memory subsystem \cite{memorganization}, \cite{Kevin:Latency}. \label{fig:memoryorganization}}
\end{figure*}

Fig. \ref{fig:memoryorganization} illustrates the organization of a modern DRAM system, which maintains a hierarchy of channel, rank, bank, DRAM chips, DRAM cells, and memory controller. Depending on the system requirement, different electronic systems can have DRAM modules of different sizes. A DRAM module is divided into one or multiple ranks. The rank is accessed in each reading/writing attempt. Rank, again, consists of several DRAM chips and provides a wide databus together. The same databus is shared among the ranks. A chip select pin is used to choose a particular rank. The width of the databus is usually $64$ bits and distributed equally among the chips inside a rank. Each DRAM chip consists of multiple banks to support the parallelism. In a memory bank, the DRAM cells are arranged in a two-dimensional array. The rows and columns of a DRAM are known as \textit{wordline} and \textit{bitline}, respectively. The row of a DRAM is also known as the page. The \textit{bitlines} are connected to the \textit{row-buffer} (a row of \textit{sense-amplifiers}). When a DRAM is read, the \textit{sense-amplifier} senses the stored charge of each memory cell and latches it to a corresponding value (`1' or `0'). A DRAM cell, the smallest unit, is used to store a single bit (`1' or `0'). The DRAM cell consists of two components: a capacitor to hold the charge and an access transistor to access the capacitor. The charging state of the capacitor determines the state of the value (`1' or `0'). A fully charged capacitor is represented by logic `1'. On the other hand, logic `0' is the representation of a capacitor with no charge. 

\renewcommand\thesubfigure{\roman{subfigure}}
\renewcommand\thesubfigure{\alph{subfigure}}


\subsection{DRAM Operation} \label{subsec:DRAMoperation}
\subsubsection{\textbf{READ Operation:}} Fig. \ref{fig:DRAMsig} presents a simplified DRAM read operation, which consists of several states. In the \textit{precharge} state, the memory controller generates a \textit{precharge} command (\textit{PRE}) to precharge all \textit{bitlines} to $V_{dd}/2$ (green line). This command also deactivates previously activated \textit{wordline}. In the next state (i.e., the \textit{activation} state), the \textit{ACTIVATE} command (\textit{ACT}) from the memory controller activates the target \textit{wordline} by raising the value of \textit{wordline} to $V_{dd}$ (violet line). Once the pass-transistor (connected to the \textit{wordline}) is ON, the charge flows from the capacitor (red line) to the attached \textit{bitline} if the stored value is `1', and moves from \textit{bitline} to the capacitor if the stored value is `0'. In the final stage, the differential \textit{sense-amplifier} senses the voltage perturbation on the \textit{bitline} and amplifies the \textit{bitline} voltage to a strong logic `1' (or `0'). Then, the \textit{sense-amplifier} latches the logic value from the \textit{bitline}. In the DRAM system, the read operation is destructive; therefore, rewriting after reading is mandatory.

\subsubsection{\textbf{WRITE Operation:}} In the \textit{write} operation, initially all \textit{bitlines} are precharged to $V_{dd}/2$ with the \textit{PRE} command. The \textit{ACT} command is applied to write data into a specific \textit{wordline}. The \textit{sense-amplifier} with desired logic value enables the corresponding \textit{bitline} to charge or discharge the connected cell capacitor. After each successful \textit{READ/WRITE} operation, the \textit{bitlines} must be precharged back to $V_{dd}/2$ to access a new set of memory cells from a different \textit{wordline}.

\subsection{DRAM Timing} \label{subsec:timing}
\begin{figure*}[ht!]
\centering
\begin{subfigure}[t!]{0.49\textwidth}
\includegraphics[width=\textwidth]{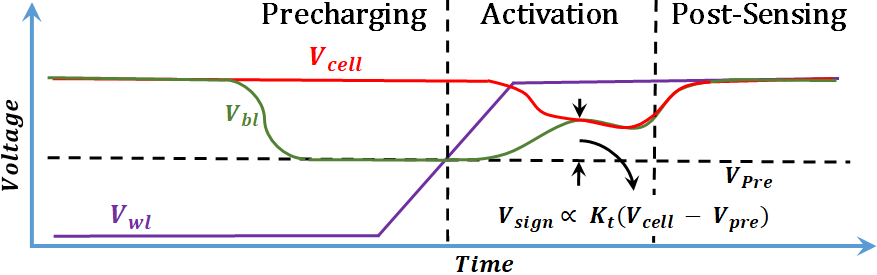} 
\caption{Signal waveform at the reading cycle.\cite{Lee:Latency}}
\label{fig:DRAMsig}
\end{subfigure}
~
\begin{subfigure}[t!]{0.49\textwidth}
\includegraphics[width=\textwidth]{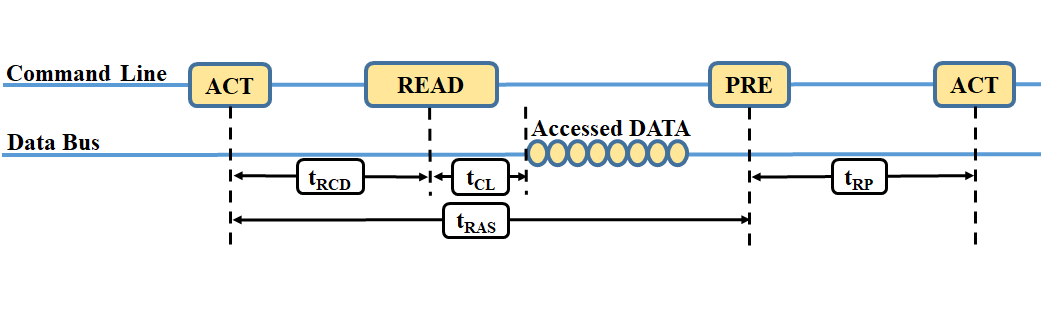}
\caption{DRAM timing at the reading cycle.\cite{Kevin:Latency}.}
\label{fig:DRAMtiming}
\end{subfigure}
\captionsetup{font={sf,small},labelfont={bf,color=accessblue}}
\caption{DRAM operation and timing. \label{fig:memoryTiming}}
\end{figure*}

Timing is critical for reliable DRAM operation. All major timing parameters of a DRAM module are presented in Fig. \ref{fig:DRAMtiming}. Initially, all \textit{bitlines} are precharged to $V_{dd}/2$. To access the data from a specific \textit{wordline}, \textit{ACTIVATE} (\textit{ACT}) command is applied to the corresponding \textit{wordline}. Once that is completed, a \textit{READ/WRITE} command is sent from the memory controller to sense the voltage perturbation on \textit{bitlines} or to write a data to the memory cells. The minimum required time interval between \textit{ACT} command and \textit{READ/WRITE} command is defined as the \textit{activation time}, $t_{RCD}$. The \textit{Column Access Strobe} (\textit{CAS}) latency $t_{CL}$ is the minimum waiting time to get the first data bit on data bus after sending a \textit{READ} command. After a successful \textit{READ/WRITE} operation, \textit{precharge} command (\textit{PRE}) is applied to deactivate the previously activated \textit{wordline} (if any) and precharge the \textit{bitlines} to its initial precharge state (i.e., to $V_{dd}/2$). If the \textit{WRITE} command is applied, the \textit{PRE} command should be further delayed by $t_{WR}$ period (\textit{write recovery} time) at the end the write data burst. The \textit{PRE} command is applied for at least $t_{RP}$ (\textit{precharge} time) duration before sending the next \textit{ACT} command. The duration between the \textit{activation} state to the beginning of the \textit{precharge} state is called \textit{row active} time or \textit{restoration} latency ($t_{RAS}$). The \textit{$t_{RAS}$ + $t_{RP}$} is the total time required to access a single row of a bank and is known as \textit{row cycle} time ($t_{RC}$). Usually, the $t_{RC}$ is in the order of 50ns for most modern DDR3 DRAMs.

\renewcommand\thesubfigure{\roman{subfigure}}
\renewcommand\thesubfigure{\alph{subfigure}}

\subsection{Existing DRAM-based PUFs} 

\subsubsection {\textbf{Retention-based DRAM PUFs (DPUFs):}}

Signatures are generated by disabling the refresh interval for a certain and sufficient amount of time \cite{retentionPUF,DRAMLatencyPUF}. The DRAM cells are leaky, and therefore, the DRAM contents need to be refreshed periodically, usually 64ms or 32ms according to the JEDEC specification \cite{jedecSTD}, to ensure the data integrity \cite{retentionPUF}.Failing to refresh periodically within this time interval introduces errors due to the leaky property of DRAM cells. The error pattern generated from the retention failure is unique from chip to chip and is used to generate device signatures \cite{retentionPUF,DRAMLatencyPUF}.

The retention-based device signature is promising but suffers from several drawbacks that hinder its use in real applications. First, the periodic refresh operation in most DRAM modules is handled internally by a memory controller. There is no efficient way to control this refresh time for an arbitrarily small region of DRAM module since the granularity for such refresh operation is predefined by the vendors. Some common control signals control memory cells under the same granularity, and therefore changing a timing parameter on one cell affects all other cells as well. On the other hand, two rows from two different granular regions can be accessed independently (but not simultaneously as they may share the same channel). For a retention-based DRAM PUF, an authentication key of sufficient length can be generated by retention failure from a small portion of a DRAM module, but the whole operation may cause unwanted data corruption of other memory cells under the same granularity \cite{jedecSTD}. Second, a key of sufficient length requires an adequate number of errors; therefore, it might need a long waiting time (order of minutes) to generate a key with desired length and quality \cite{DRAMLatencyPUF}. Third, the retention time is heavily temperature dependent, which makes the key sensitive to temperature variations \cite{retentionPUF, retentionPUF2, retentionPUF3, retBehavior, Hassan:ChargeCache, leakagedistribution}. Previous studies show that the \textit{bit error rate (BER)} increases exponentially with the temperature; the key generation scheme requires a longer time interval between two refresh operations at a lower temperature \cite{systemHeat}.
The required time to generate the key is also a function of the size of the memory segment. A smaller segment requires longer evaluation time than a larger one \cite{DRAMLatencyPUF}. Therefore, the designer must decide on area vs. time overhead. Several techniques can be used to address the above challenges but with a limited gain \cite{systemHeat,partialRestore,reducedvoltageOP,DRAMStorageProcessVar,retentionPUF,parbor, Wang:Look-Ahead, Patel:REAPER}. 

\subsubsection {\textbf{Latency-based DPUFs:}}
The reduction in $t_{RCD}$ introduces erroneous read/write operation (see Section \ref{subsec:timing}), which can be used to generate device signatures \cite{DRAMLatencyPUF}. This latency-based PUF generates signature at a much faster rate \cite{DRAMLatencyPUF}. The reported result shows that the mean evaluation time is $\sim$88.2ms (outperforms all previously proposed retention-based DPUFs \cite{retentionPUF,retentionPUF2,retentionPUF3}). However, it still requires multiple row cycles to evaluate the PUF response. This latency-based DPUF also needs a filtering mechanism in each access that adds both hardware and computational overheads.

\subsubsection {\textbf{Start-up based DPUFs:}}
In start-up based DPUF\cite{DPUFstartup}, the device signature is generated from the start-up states of DRAM cells. Initially, the bitlines are charged to $V_{dd}/2$. But the process variations on the storage capacitor slightly deviate the \textit{bitline} voltage to $V_{dd}/2+\delta$ or $V_{dd}/2-\delta$, where $\delta$ represents a small voltage. The \textit{sense-amplifier} senses the voltage difference to `1' or `0', accordingly. Upon power-up, the DRAM cells generate `1's and `0's randomly. The significant challenges of a start-up based DPUF are: (i) requirement of a power-cycle and (ii) a time gap between the turn-OFF and turn-ON is required to avoid a strong correlation between the data before turn-OFF and the signature.

\subsubsection {\textbf{Rowhammer DPUFs:}}
The errors caused by the rowhammer disturbance are used to generate device signatures \cite{Schaller:hammerPUF, Anagnostopoulos:hammerPUF}. This technique does not require any additional power cycle. However, the average evaluation time of a rowhammer PUF is in order of minutes and therefore might not be suitable in many applications. Besides, all DRAMs are not vulnerable to rowhammer \cite{Schaller:hammerPUF}.

\subsection{Motivations} \label{Motivations}
Below, we summarize the major motivations of our proposed work. 
\begin{itemize}
\item \textbf{Waste of DRAM Power Cycle:} Start-up based key generation requires a DRAM power cycle to obtain device signatures \cite{DPUFstartup}. Hence, the whole system needs a power cycle (i.e., a turn-off and a turn-on) to obtain the PUF response. Therefore, this type of PUF cannot be evaluated while the system is in operation.
\item \textbf{Large Evaluation Time:} Rowhammer-based and retention-based key generation techniques require an order of minutes to generate enough bit failures and therefore not suitable for many applications \cite{retentionPUF,retentionPUF2,retentionPUF3, AVATAR, Schaller:hammerPUF, Anagnostopoulos:hammerPUF}. On the other hand, the existing latency-based DPUF still needs multiple row cycles (reading one data burst at each cycle) to evaluate the PUF key \cite{DRAMLatencyPUF} since the reduction in activation time only affects the first few bits in the cache line (see Section \ref{subsec:timing}).
\item \textbf{Destructive:} Retention-based key generation is destructive. The DRAM granularity causes random failed bit throughout the smallest granular region (usually a rank). Note that the DRAM refresh can be disabled only at the granularity of channels \cite{jedecSTD}. A dedicated memory might need to be used to overcome this problem but at the expense of additional hardware. The start-up based and rowhammer-based DPUFs are also destructive.
\item \textbf{Disruptive:} DRAM granularity keeps the entire DRAM rank busy during each access. Hence, such kind of PUF evaluation blocks the access on the target DRAM region by other applications for a long time. Though the existing latency-based DRAM PUF \cite{DRAMLatencyPUF} solves the problem of long evaluation time and unwanted data failure (due to the granularity), it still needs a filtering mechanism to evaluate PUF in each access, which introduces additional computational and timing overheads.
\end{itemize}

\section{PreLatPUF: Prechareg Latency-based PUF} \label{sec:designoverview} 
In this section, we present the proposed PreLatPUF, cell characterization, and cell selection algorithm.

\subsection{Precharge Latency and Source of Variations} \label{sec:divSig}

The latency is defined as the time required to move charge during read/write operation. In modern DRAM architecture, multiple DRAM cells are connected to the same \textit{bitline} through access transistors. The DRAM vendor provides the minimum required timing latency to perform a reliable read/write operation. Erroneous read/write operation is observed if the minimum timing latency is not maintained \cite{Kevin:Latency}. In our experimental results, the following observations have been discovered that are also consistent with \cite{Kevin:Latency} and \cite{ Kim:SolarDRAM}.
\begin{itemize}
\item \textbf{Observation 1}: A reduced $t_{RCD}$ only affects the first accessed column/cache line.
\item \textbf{Observation 2}: A reduced $t_{RP}$ might affect almost all cells of a row.
\item \textbf{Observation 3}: Almost no bit error is introduced at the reduced $t_{RAS}$.
\end{itemize}

From the above observations, we can conclude that the reduction in $t_{RCD}$ or $t_{RP}$ can be used to generate device signatures from a DRAM. The $t_{RCD}$-based PUF has been proposed in \cite{DRAMLatencyPUF} that needs an additional filtering mechanism and several row cycles (discussed in Section \ref{Motivations}). In this article, we use the $t_{RP}$ variations to generate device signatures. 

\begin{figure}[ht!]
\centering
\captionsetup{justification=centering, margin=0.5cm}
\includegraphics[width=0.48\textwidth]{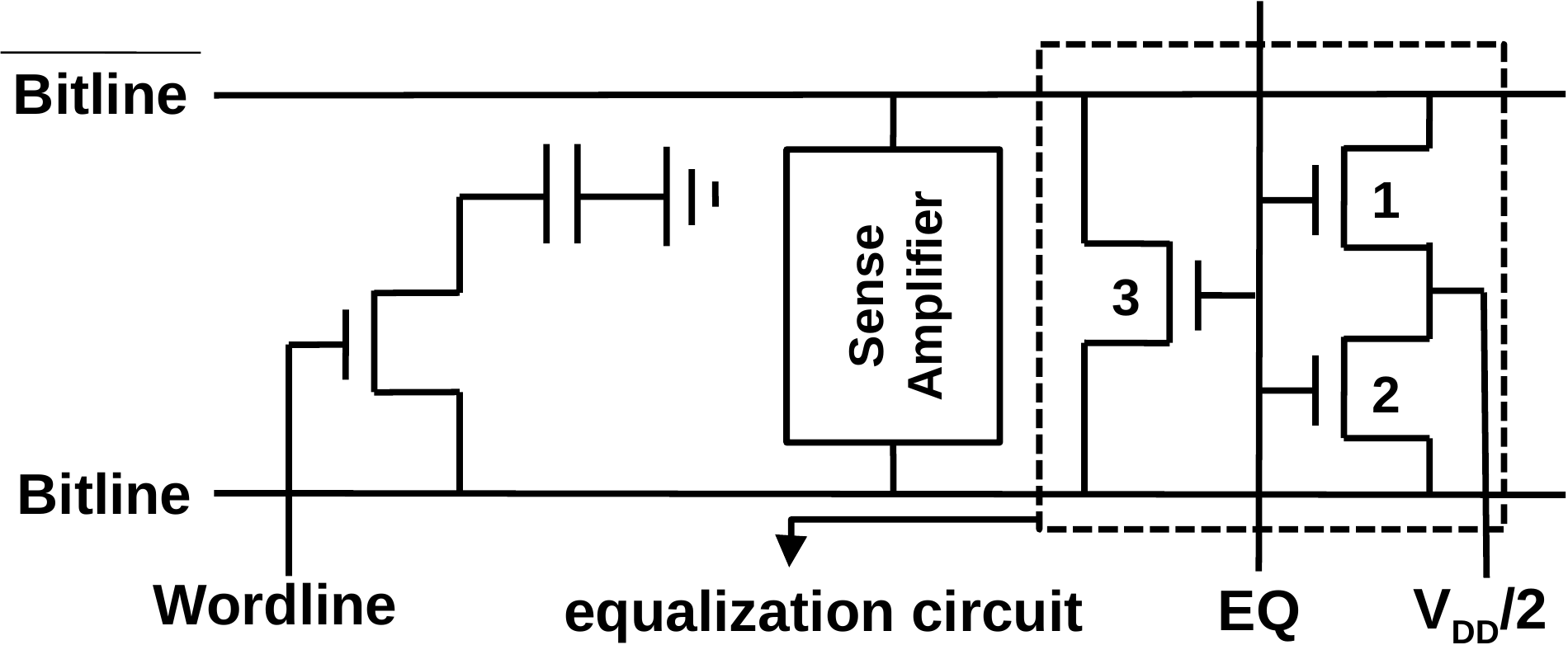}
\captionsetup{font={sf,small},labelfont={bf,color=accessblue}}
\caption{Simplified structure of precharge circuit.}
\label{fig:prechargeckt}
\end{figure}

 The DRAM cell characteristics at the reduced $t_{RP}$ mostly rely on the internal structure of a DRAM module, process variations, layout variations, data dependency, etc. \cite{softMC, Khan:dataDependency, designInducedTimingVar, AVATAR, parbor, Kevin:Latency, DRAMLatencyPUF, btalukeder:DRNG}. Fig. \ref{fig:prechargeckt} presents a simplified structure of the DRAM precharge circuit \cite{Jacob:DRAM}. In a DRAM module, each DRAM cell is connected to a \textit{bitline} through an access transistor and each \textit{bitline} has a corresponding $\overline{bitline}$ that provides the complementary data
(see Fig. \ref{fig:prechargeckt}). Each \textit{bitline} and $\overline{bitline}$ pair contain a \textit{sense-amplifier} and an equalization circuit. At the precharge state, the transistor 1 and 2 of the equalization circuit create a conducting path with a voltage source $V_{DD}/2$. On the other hand, the transistor 3 of the equalization circuit creates a conducting path between \textit{bitilne} and $\overline{bitline}$. With the proper precharge time, the transistor 1 and 2 get enough time to precharge the \textit{bitline} pair to $V_{DD}/2$, and the transistor 3 further ensures the equalization of voltage on the \textit{bitline} pair. After turning ON the access transistor, the \textit{bitline} voltage is perturbed by the stored charge in the capacitor. Then the perturbed voltage is sensed and amplified with the \textit{sense-amplifier}. However, at the reduced precharge time, the transistor 1 and transistor 2 might not get enough time to precharge the \textit{bitline} pair equally to $V_{DD}/2$. Therefore, the \textit{bitline} and the $\overline{bitline}$ might deviate from VDD/2.
The variations on \textit{RC} path delay and the capacitance of the \textit{bitline} follow the Gaussian distribution \cite{timingProbability, timingProbability2, capProbability}, and two different DRAM cells of same physical length may have different $t_{RP}$s.

In addition to this, the process variation also might introduce slight variation on the charge storage capacity of the the DRAM cells. Hence, during the READ operation, the intensity of the voltage perturbation on a \textit{bitline} might vary from one memory cell to another memory cell \cite{Shin:NUAT}.
As a result, these DRAM cells may behave differently at the reduced \textit{precharge} time \cite{Lee:Latency}. In addition, different vendors may follow different kind of configurations (e.g., \textit{open bitline} array structure, \textit{folded bitline} array structure etc. \cite{Jacob:DRAM}), which may lead to different faulty outputs at the reduced \textit{$t_{RP}$}.

Note that the minimum value of $t_{RP}$ is required to deactivate the previously activated row to avoid correlation among the outputs and the contents of the previously activated row. The minimum value of \textit{$t_{RP}$} is determined empirically, and it may vary from module to module (discussed in Section \ref{sec:cellChar}).

\subsection{\textbf{Characterization}}
We characterize the DRAM cells to understand the data dependency, spatial correlation, etc. in order to obtain robust PUF signatures. The characterization phase is conducted by observing the outputs with different types of input patterns (e.g., all 1's, all 0's or checkerboard pattern). The term `input value' or `input pattern' is used for the pattern that is written in the DRAM memory module with standard timing parameters. On the other hand, the `output pattern' refers to the output that is read back at the reduced ${t_{RP}}$. A particular input pattern is applied several times (more on Section \ref{sec:results}) to study the temporal variation (i.e., measurement variation). Based on the data correctness (or incorrect/faulty behavior), we divide the DRAM cells into two major categories:
\begin{itemize}
\item \textbf{Non-faulty Cells:} These memory cells do not show any errors at the reduced ${t_{RP}}$ and retain correct data regardless of the input data pattern.
\item \textbf{Incorrect/Corrupted/Faulty Cells:} These memory cells fail to output the original data (i.e., the input pattern and output pattern are different). The errors might be independent or dependent on the input data. 
\end{itemize}

Based on the temporal variations, again, we categorize the incorrect/faulty cells into the following types: 
\begin{itemize}
\item \textbf{Noisy Cells:} Error pattern varies from measurement to measurement because of internal/external noise for these types of cells. Some of these cells can be useful to generate random number \cite{btalukeder:DRNG}. Some of these cells can be used to create PUF but might require a large ECC \cite{hassRahman}.

\item \textbf{Robust/Measurement-invariant Cells:} These cells do not show any temporal variation, i.e., cell outputs are independent of measurements. These cells are tolerant to internal and external noise and ideal for PUF.
\end{itemize}

In addition, the outputs at the reduced $t_{RP}$ might depend on the memory cell contents (i.e., written patterns) due to the coupling effect of neighborhood cells \cite{coupling, Khan:dataDependency}. Based on the data dependency, we categorize the DRAM cells into following types:
\begin{itemize}
\item \textbf{Pattern Independent Cells:} These types of cells exhibit the same output (at the reduced $t_{RP}$) regardless the input patterns. The experimental results show that (details in Section \ref{sec:results}), most of the DRAM cells from the major vendors are \textit{pattern independent}. In this paper, we have only focused on the \textit{`pattern independent'} cells for PUF implementation.
 
\item \textbf{Pattern Dependent Cells:} The output patterns for these cells are different for different input patterns. Therefore, these cells can be the ideal candidates for the PUF that possesses enhanced challenge-response pair \cite{strongDRAM1, Ganta:PUF, Maiti:PUF}.

\end{itemize}

\subsection{\textbf{Cell Selection Algorithm}} \label{sec:cellSelection}
In this paper, we only focus on the \textit{pattern independent} cells. 
The experimental results show that some of the \textit{pattern independent} cells are \textit{strong `1'} and some of them are \textit{strong `0'}. Besides the reproducibility, it is important that the generated key is random and unique as well. Entropy is used to measure the randomness (i.e., the unpredictability) of a bitstream \cite{hassRahman, TRNG}. A binary string of randomly distributed 0's and 1's with equal probability possess high entropy \cite{hassRahman, TRNG, entropy}. Not all cells can be used to generate PUF because some DRAM cells create deterministic outputs. We scan each row to find the most suitable cells for generating robust and random keys. We observe that the generated outputs using all \textit{pattern independent} bits of every word (a word is 64 bits wide) suffer from poor entropy. As a part of the entropy test, we count the ratio between the occurrence of 1's and the occurrence of 0's. Our objective is to generate a key that has an equal number of 1's and 0's. The raw outputs show that there is a considerable imbalance between the number of 0's and 1's if we count each failed bit from all words. Therefore, all bits of every word are not suitable for key generation. It is observed that some specific bits of every word of a row produce a predictable outcome. For example, for a particular memory bank, the first bit of every word of a specific row is always read as `0' at a reduced $t_{RP}$. The binary string ($V_1$) formed with the first bits of the words cannot be used to generate keys since the \textit{Hamming weight}\footnote{The \textit{Hamming weight} is defined as the total number 1's (or 0's) in a bitstream.} of the $V_1$ is 0\%. The explanation of this phenomenon as follows: a 64-bit DRAM module is analogous to a combination of 64 2-D memory arrays (distributed into multiple DRAM chips), and each memory array contributes to every word by providing one bit. For example, the $5^{th}$ memory array is responsible for the $5^{th}$ bit of the word. The impact of reduced $t_{RP}$ may vary among memory arrays. In our proposed bit selection algorithm, we use an important metric: \textit{Hamming weight}. A 50\% of \textit{Hamming weight}, which is ideal for a key, means that the binary string has an equal number of 1's and 0's. Similar to $V_1$, we create a binary string $V_2$ with the second bit of each word in a row. Similarly, the binary string generated from the $i^{th}$ bit of each word is $V_i$. The $i^{th}$ bit of the word is considered as the \textit{eligible bit} if it produces a random binary string $V_i$ with a $ \sim 50\%$ Hamming weight.

To get the most suitable cells for robust PUF, we propose an algorithm (Algorithm \ref{selectionAlgo}) for selecting the qualified memory cells and their locations. In practice, not all binary strings in $\mathcal{V}= \{V_1, V_2, ..., V_{64}\}$ experiences a 50\% of \textit{Hamming weight}. Therefore, we choose only those binary strings that fall into a range of allowable Hamming weight ($H_{min}$ to $H_{max}$). All eligible bits (of words) from a row $\mathcal{R}_x$ can be defined as Eq. \eqref{eq1}.
Table \ref{tab:cellSelect} shows a simplified explanation of selecting eligible bits, where we have presented all memory cells from an imaginary row that has 4-bit ($V_1$ to $V_4$) wide 16 words ($W_1$ to $W_{16}$). We have produced the first string $V_1$ by only taking the first bit from each word, $V_2$ by only taking the second bit from each word and so on. The rightmost column of Table \ref{tab:cellSelect} presents the Hamming weight ($HW$) of each string. For better randomness, the \textit{Hamming weight} of each string should be 50\% (8 in this case). However, the silicon results show that it is not always achievable. Therefore, we have to choose a lower limit ($H_{min}$) and an upper limit ($H_{max}$) of \textit{Hamming weight}. Let's assume, the chosen values of $H_{min}$ and $H_{max}$ are 5 and 11, respectively. As a result, only cells under the $V_2$ and $V_4$ can be used for PUF operation (as Hamming weight of $V_2$ and $V_4$ are between 5 and 11, see Table \ref{tab:cellSelect}). So, according to the Eq. \eqref{eq1}, the set of eligible bits is $\beta_{\mathcal{R}_{x}} = \{2, 4\}$.

If the row $\mathcal{R}_x$ consists of \textit{n} words, then we can create a binary string from each word by only considering the qualified bits (i.e., the cells that satisfy Eq. \eqref{eq1}). For example, if we consider the $i^{th}$ word $W_i$ from row $\mathcal{R}_x$, then, $W_i^{\beta_{\mathcal{R}_{x}}}$ is a binary string by taking bits which are the elements of $\beta_{\mathcal{R}_{x}}$. So, all allowable data bits from the $\mathcal{R}_x$ can be presented as the Eq. \eqref{eq2}. Here, $\mathcal{M}_{\mathcal{R}_{x}} $ is a single dimensional binary string containing all eligible data bits from $\mathcal{R}_x$. According to the Table \ref{tab:cellSelect}, and Eq. \eqref{eq2}, $\mathcal{M}_{\mathcal{R}_{x}} = [11, 00, 10, 10, 00, 11, 01, 00, 01, 11, 00, 01, 11, 01, 10, 01]$.
\begin{multline}\label{eq1}
\beta_{\mathcal{R}_{x}} = \left \{b\in \left\{1,2,3,..., 64\right\}: \right. \\ \left. H_{min} < Hamming\_weight\_of(V_b) < H_{max}\right \}
\end{multline}

\begin{flalign}\label{eq2}
&\ \ \ \ \mathcal{M}_{\mathcal{R}_{x}} = [W_1^{\beta_{\mathcal{R}_{x}}}, W_2^{\beta_{\mathcal{R}_{x}}}, W_3^{\beta_{\mathcal{R}_{x}}}, ..., W_n^{\beta_{\mathcal{R}_{x}}} ]&
\end{flalign}

\begin{table*}[ht!]
\captionsetup{singlelinecheck=false,font={sf,small},labelfont={bf,color=accessblue}}
\caption{Selecting appropriate cells with cell selection algorithm.}
\centering
\setlength\tabcolsep{7pt} 
\begin{tabular}{|c|c|c|c|c|c|c|c|c|c|c|c|c|c|c|c|c|c|}
\hline
\rule{0pt}{2ex} & $W_1$ & $W_2$ & $W_3$ & $W_4$ & $W_5$ & $W_6$ & $W_7$ & $W_8$ & $W_9$ & $W_{10}$ & $W_{11}$ & $W_{12}$ & $W_{13}$ & $W_{14}$ & $W_{15}$ & $W_{16}$ & $HW$ \\ \hline
\rule{0pt}{2ex} $V_1$ & 1 & 1 & 1 & 1 & 1 & 0 & 1 & 1 & 1 & 1 & 1 & 1 & 1 & 1 & 1 & 1 & 15 \\ \hline
\rule{0pt}{2ex} $V_2$ & 1 & 0 & 1 & 1 & 0 & 1 & 0 & 0 & 0 & 1 & 0 & 0 & 1 & 0 & 1 & 0 & 7 \\ \hline
\rule{0pt}{2ex} $V_3$ & 0 & 0 & 0 & 0 & 0 & 0 & 0 & 1 & 0 & 0 & 0 & 0 & 0 & 1 & 0 & 0 & 2 \\ \hline
\rule{0pt}{2ex} $V_4$ & 1 & 0 & 0 & 0 & 0 & 1 & 1 & 0 & 1 & 1 & 0 & 1 & 1 & 1 & 0 & 1 & 9 \\ \hline
\end{tabular}
\label{tab:cellSelect}
\end{table*}

\begin{algorithm}

\hspace*{\algorithmicindent} \textbf{Input:} \\
\hspace*{\algorithmicindent} \textit{mem\_data}: A $R_n\!\times\! C_n\!\times\! B_n$ matrix, containing pattern independent data. An element of \textit{mem\_data} can be empty (if the corresponding memory cell is not pattern independent) or `0' or `1'. \\
\hspace*{\algorithmicindent} \textit{$H_{min}$ \& $H_{max}$}: Minimum and maximum allowable Hamming weight as described in sec \ref{sec:cellSelection}. \\
\hspace*{\algorithmicindent} \textbf{Output:} \\
\hspace*{\algorithmicindent} $\mathcal{R}$: 1D array, contains the list of qualified rows which holds \textit{qualified bits} for PUF generation \\ 
\hspace*{\algorithmicindent} $\beta$: 2D array, $i^{th}$ row is associated with the $i^{th}$ row of $\mathcal{R}$. Each row of $\beta$ contains all \textit{qualified bits} from each word of the corresponding row.  \\

\begin{algorithmic} [1]
\caption{Selecting qualified memory cells.}
\label{selectionAlgo}
\STATE ${\beta = \ [\ ];}$ // $\beta\ initialized\ with\ empty\ matrix$
\STATE ${\mathcal{R} = \ [\ ];}$ // $\mathcal{R}\ initialized\ with\ empty\ matrix$
\STATE ${bit\_count = 0;}$
\STATE ${row\_count = 0;}$
\STATE ${row\_flag = false;}$
\FOR {$r = 1$ to $R_n$}
\FOR {$b = 1$ to $B_n$}
\STATE ${V_b = \ [\ ];}$
\STATE ${k = 0;}$
\FOR {$i = 1$ to $C_n$}
\STATE ${temp = mem\_data\left(r,i,b\right);}$
\IF {$is\_pattern\_independent(temp) == true$}
\STATE ${V_b\left(k\right) = temp;}$
\STATE ${k = ++;}$
\ENDIF
\ENDFOR
\STATE ${h = Hamming\_weight\_of \left(V_b\right);}$
\IF {$h > H_{min}$ \&\& $h < H_{max}$}
\STATE ${row\_flag = true;}$
\STATE ${\beta\left(row\_count, bit\_count\right) = b;}$
\STATE ${bit\_count ++;}$
\ENDIF
\ENDFOR
\IF {$row\_flag == true$}
\STATE ${\mathcal{R}\left(row\_count \right) = r;}$ 
\STATE ${row\_count ++;}$
\STATE ${bit\_count = 0;}$
\STATE ${row\_flag = false;}$
\ENDIF
\ENDFOR
\end{algorithmic}
\end{algorithm}

However, the length of the key can be larger than the number of qualified memory cells in a binary string $\mathcal{M}_{\mathcal{R}_{x}}$. In this case, we will have to use more than one binary string from the multiple rows. Algorithm \ref{selectionAlgo} is designed to select the \textit{qualified bits} (i.e., the cells that satisfy Eq. \eqref{eq1}) from each row. From now on to the rest of our discussion, the $b^{th}$ bit of the 64-bit data word, accessed from the location \textit{(r,c)}, will be noted as \textit{(r,c,b)} where, \textit{r} is the row number (or page number), and \textit{c} is the column number ($c^{th}$ word of the row \textit{r}). In Algorithm \ref{selectionAlgo}, \textit{$R_n$}, \textit{$C_n$}, and \textit{$B_n$} are the total number of rows, total number of columns, and the word width respectively (constant for a specific memory module). In our experiment, we have used 1GB memory modules, where, \textit{$R_n$ = 16384}, \textit{$C_n$ = 1024}, and \textit{$B_n$ = 64}).

\begin{figure}[ht!]
\centering
\captionsetup{font={sf,small},labelfont={bf,color=accessblue},justification=centering, margin=0cm}
\includegraphics[width=0.25\textwidth]{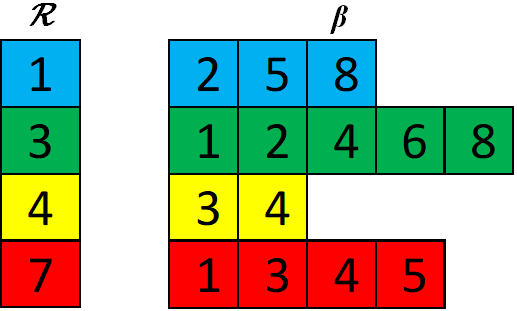} 
\caption{Qualified row position and corresponding bit position in words.}
\label{fig:qualifiedCells}
\end{figure}

In the proposed Algorithm \ref{selectionAlgo}, a one-dimensional array $\mathcal{R}$ and a two-dimensional array $\beta$ together hold the memory locations of the qualified DRAM cells. The $\mathcal{R}$ holds all eligible row (or page) addresses and $\beta$ holds corresponding \textit{qualified bit} number of the row. For example, $\mathcal{R} = {1, 3, 4, 7}$ represents that $1^{st}$, $3^{rd}$, $4^{th}$, and $7^{th}$ rows (or pages) are marked as the qualified rows (see Fig. \ref{fig:qualifiedCells}). $\beta$ (on right side) of the fig. \ref{fig:qualifiedCells} represents corresponding locations of the eligible bits. For example, for $\mathcal{R}=1$, the $\beta = \{2,5,8\}$. i.e. $2^{nd}$, $5^{th}$ and $8^{th}$ bit of all words from row 1 can be used to generate key.

\subsection{\textbf{Registration}}
In the \textit{registration} phase, we generate a golden data set (i.e. challenge-response data set) using Algorithm \ref{goldenDataAlgo}, which can be used to generate robust signatures. We assume that the golden data set is created and stored in a trusted environment. In the Algorithm \ref{goldenDataAlgo}, we use qualified memory cells that are obtained using Algorithm \ref{selectionAlgo}. In Algorithm \ref{goldenDataAlgo}, the \textit{goldenDataLoc} holds the logical locations of eligible memory cells and the \textit{goldenData} saves the outputs that are accessed from the corresponding locations at the reduced \textit{$t_{RP}$}. The \textit{goldenDataLoc}, \textit{goldenData}, and the reduced value of \textit{$t_{RP}$} will be used as the golden data set for future authentication.

\begin{algorithm}

\hspace*{\algorithmicindent} \textbf{Input:} \\
\hspace*{\algorithmicindent} \textit{mem\_data}: A $R_n\!\times\! C_n\!\times\! B_n$ matrix, containing pattern independent data. An element of \textit{mem\_data} can be empty (if the corresponding memory cell is not pattern independent) or `0' or `1' . \\
\hspace*{\algorithmicindent} \textit{$\beta$ \& $\mathcal{R}$}: generated from algorithm \ref{selectionAlgo}. \\
\hspace*{\algorithmicindent} \textbf{Output:} \\
\hspace*{\algorithmicindent} \textit{goldenDataLoc}: A boolean matrix of size $R_n\!\times\! C_n\!\times\! B_n$. \textit{goldenDataLoc(r,c,b)} is \textit{true} if corresponding memory cell qualified for the PUF application \\ 
\hspace*{\algorithmicindent} \textit{goldenData}: Matrix of size $R_n\!\times\! C_n\!\times\! B_n$, contains pattern independent output of those memory cells that are marked as \textit{true} in \textit{goldenDataLoc} matrix.\\

\begin{algorithmic} [1]
\caption{Generating golden data.}
\label{goldenDataAlgo}
\STATE ${goldenDataLoc = boolean\_matrix\left(Rn,Cn,Bn\right);}$
\STATE ${goldenData = matrix\left(Rn,Cn,Bn\right);}$
\FOR {$i = 1$ to $length(\mathcal{R})$}
\FOR {$j = 1$ to $length(\beta \left(i,1  \ to \ end \right ))$}
\FOR {$k = 1$ to $C_n$}
\STATE ${temp = mem\_data\left(\mathcal{R}\left(i\right),k,\beta\left(i,j\right)\right);}$
\IF {$is\_pattern\_independent(temp) == true$}
\STATE ${goldenDataLoc\left(\mathcal{R}\left(i\right),k,\beta\left(i,j\right)\right) = true;}$
\STATE ${goldenData\left(\mathcal{R}\left(i\right),k,\beta\left(i,j\right)\right) = temp}$;
\ENDIF
\ENDFOR
\ENDFOR
\ENDFOR
\end{algorithmic}
\end{algorithm}

\section{Result and Analysis} \label{sec:results}
Our results are based on experiments conducted with six memory banks from two commercial DDR3 memory modules of two major memory vendors\footnote{vendor \textit{A}: \textit{Micron}, vendor \textit{B}: \textit{Samsung}} (namely \textit{A} and \textit{B}). We used \textit{SoftMC} (Soft Memory Controller \cite{softMC}) along with the \textit{Xilinx ML605} Evaluation Kit which is embedded with \textit{Virtex-6} FPGA. \textit{SoftMC} uses \textit{Riffa} \cite{riffa} framework to establish communication between a host PC and the evaluation board through x8 PCIe bus. To check the design reliability against voltage variation, we used a USB interface adapter evaluation module \cite{USB2GPIO} for controlling the voltage of the memory module very precisely. 

The experiment was performed in two steps. First, an 8-bit pattern was written at the regular timing parameter and then read it back at the reduced timing parameter. The reading operation was done in a single row cycle, i.e., we activated one \textit{wordline} at a time and then read all \textit{bitlines} with consecutive burst. Here, each data burst was able to capture the data from successive 8 \textit{bitlines}. This whole process was done at the nominal operating voltage and room temperature (i.e., 25$\degree$C and 1.5V for all modules). To obtain and analyze the error pattern, we first checked the \textit{Hamming Distance} between the written pattern (input pattern) and the pattern that was read out (output pattern) with the reduced timing parameter. Then, failed bits were analyzed for additional information (e.g., spatial distribution, pattern dependency, etc.). Four sets of 8-bit input patterns (\textit{0xFF}, \textit{0xAA}, \textit{0x55}, \textit{0x00}) were used to characterize the DRAM cells. For each set of the input pattern, we repeated our experiment five times (hence, produced 20 sets of data) to study the temporal variation. Independent analysis is done by choosing random memory banks (four from vendor \textit{A} and two from \textit{B}; each consists 128MB memory cells).

We conducted our experiment on DRAM memory module by changing the \textit{activation} time (\textit{$t_{RCD}$}), \textit{restoration} time (\textit{$t_{RAS}$}), and \textit{precharge} time (\textit{$t_{RP}$}).

\subsection{\textbf{Reduced Latency: Activation Time vs. Precharge Time}} \label{actVSpre}

We read a whole row in a single row-cycle to evaluate the error patterns generated at the reduced \textit{$t_{RCD}$}. Two 32-byte (double-data rate) memory chunks were read with each burst (with 8-bit burst length, i.e., eight words can be accessed at a time while each word corresponds to 64-bit data). From now on to rest of our discussion, we will use the notation \textit{$t_{A,x}$} to present the reduced timing parameter \textit{$t_A$}, where $x$ is the reduced value of the timing parameter in nanosecond. At a reduced activation time (e.g., at \textit{$t_{RCD,5.0}$}), failed bits were only observed at the first accessed cache line (i.e., in the first 64-byte data) and therefore it needs several read cycles.
All memory banks from our selected manufacturers exhibit similar characteristics. Such behavior is observed because the target \textit{wordline} is fully activated before accessing the second content of the cache line (see appendix \hyperref[app:A]{A}). Note that \cite{Kevin:Latency} and \cite{DRAMLatencyPUF} also presented similar observation. 

\begin{figure}[ht!]
\centering
\captionsetup{font={sf,small},labelfont={bf,color=accessblue},justification=centering, margin=0.5cm}
\includegraphics[width=0.48\textwidth]{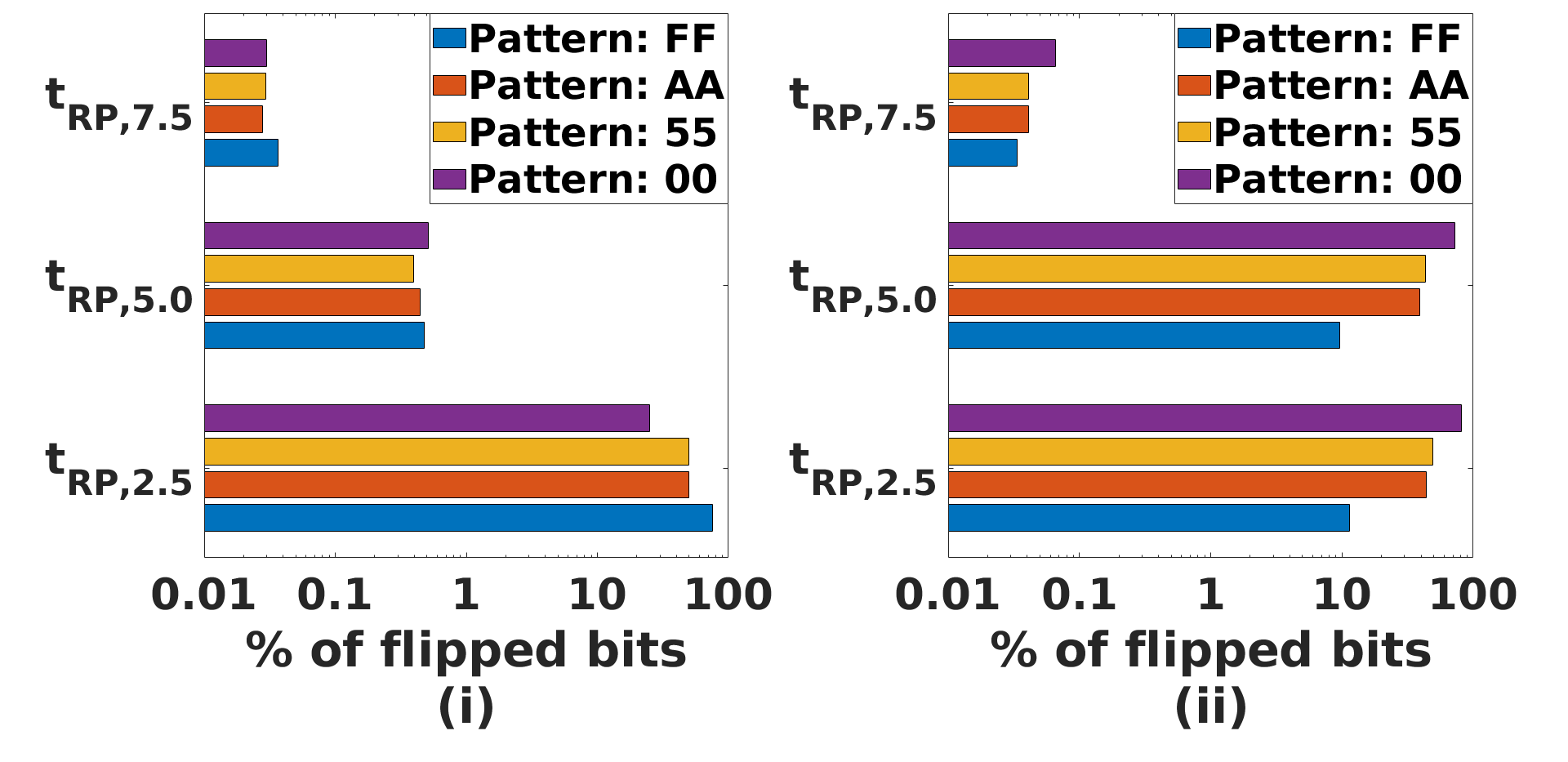} 
\caption{ \textit{$t_{RP}$} vs. \% of failed bits- (i) from vendor \textit{A} and (ii) from vendor \textit{B} (the horizontal axis shown in logarithmic scale).}
\label{fig:errorVStRP}
\end{figure}

On the other hand, the experimental results show that enough reduction in \textit{$t_{RP}$} creates errors uniformly across the whole word. In addition, it requires only a single row-cycle. Fig. \ref{fig:errorVStRP} shows that the percentage of failed bits in two random banks from two vendors for different input patterns at the reduced \textit{$t_{RP}$}. We observed the first error(s) at \textit{$t_{RP,7.5}$}. We reduced the \textit{$t_{RP}$} to \textit{$t_{RP,7.5}$}, \textit{$t_{RP,5.0}$}, and \textit{$t_{RP,2.5}$} to observe the behavior of failed bits. The results show that the total number of failed bits are $\ll 1\%$ at \textit{$t_{RP,5.0}$} for vendor \textit{A}. The rate of the failed bits increases at a much faster rate as we decrease the $t_{RP}$ further. For \textit{vendor B}, the total number of failed bits are $\ll 1\%$ at \textit{$t_{RP,7.5}$} but increase significantly at \textit{$t_{RP,5.0}$}.

We also discover that DRAMs from different manufacturers react differently for a given input pattern. Fig. \ref{fig:errorVStRP} (left) shows that most of the cells produce faulty outputs with the input pattern that has all 1's, but most of the bits are faultless when the input pattern is all 0's. On the other hand, we observe in Fig. \ref{fig:errorVStRP} (right) that most of the bits are failed when the input pattern is all 0's but most of the bits are seemed to be correct when the input pattern is all 1's. In the left figure, the number of failed bits for input pattern \textit{0xFF} is higher because the pattern independent `0' (output always `0' regardless of the input pattern) cells are dominant for this module. In the right figure, the number of failed bits for the input pattern \textit{0x00} is higher as the pattern independent `1' (output always `1' regardless of the input pattern) cell is dominant for this module. 

We can conclude from the results that (i) reducing \textit{precharge} time is superior to the reducing \textit{activation time} for generating quality signatures in a single row-cycle, and (ii) the erroneous behavior depends on the input pattern, the DRAM architecture, process variations, the amount of reduction in $t_RP$, etc.

\subsection{\textbf{Cell Characterization}} \label{sec:cellChar}
The silicon results show that a different number of faulty outputs are generated at different reduced $t_{RP}$s. In addition, we must ensure that the reduced $t_{RP}$ is also capable of closing the previously activated row (as discussed in Sec. \ref{sec:divSig}). The chosen value of $t_{RP}$ is empirical (denoted as $t_{RP,PUF}$ for clarification), which used to characterized DRAM cells, and to evaluate the PUF responses. Note that the $t_{RP,PUF}$ might vary from module to module. Also, the cell characterization is done at nominal voltage and room temperature (i.e., 25$\degree$C and 1.5V for all modules).

\renewcommand\thesubfigure{\roman{subfigure}}
\begin{figure*}[ht!]
\centering
\begin{subfigure}[t!]{0.4\textwidth}
\includegraphics[width=0.9\textwidth]{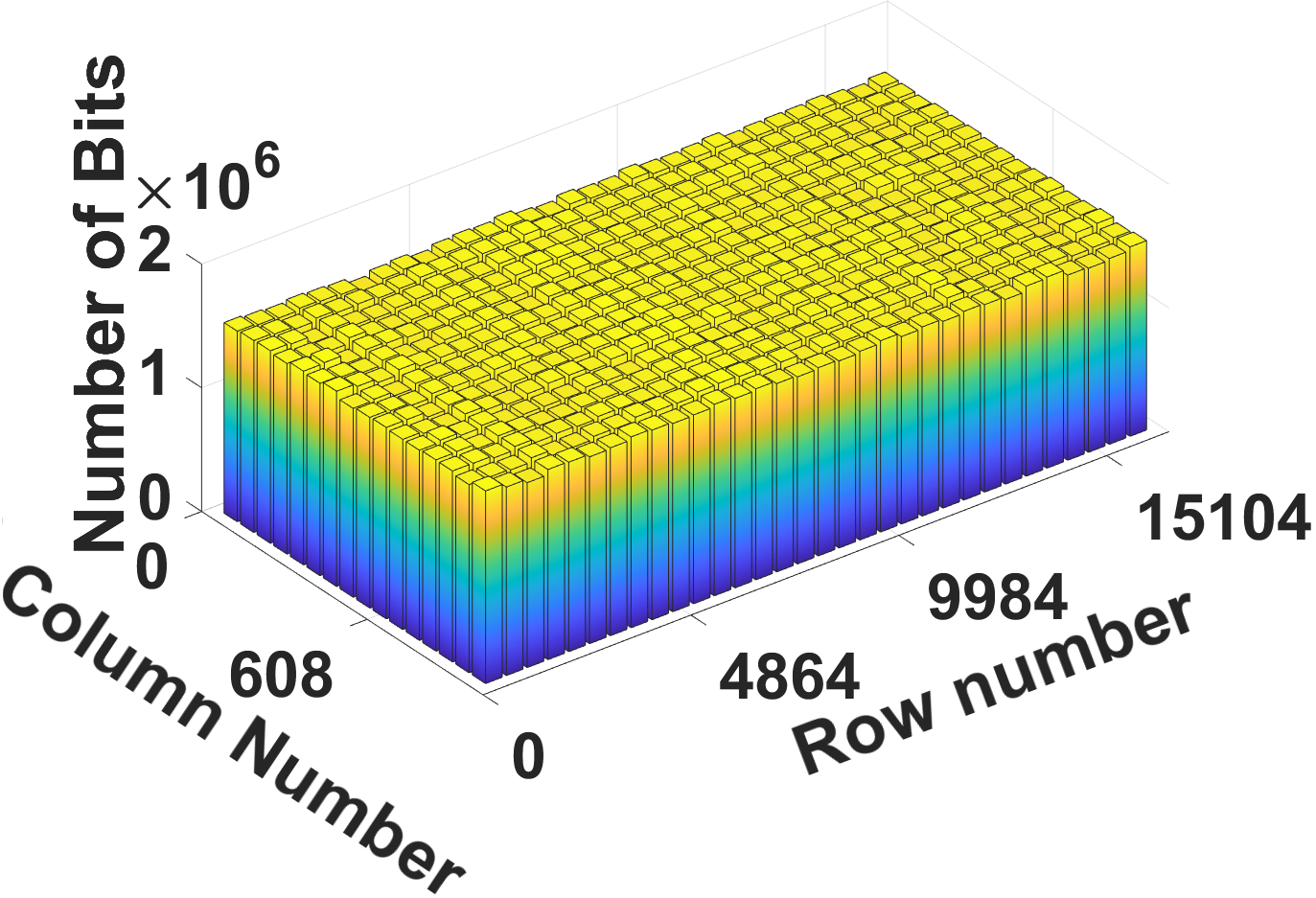} 
\caption{}
\label{fig:patIndB0}
\end{subfigure}
~
\begin{subfigure}[t!]{0.4\textwidth}
\includegraphics[width=0.9\textwidth]{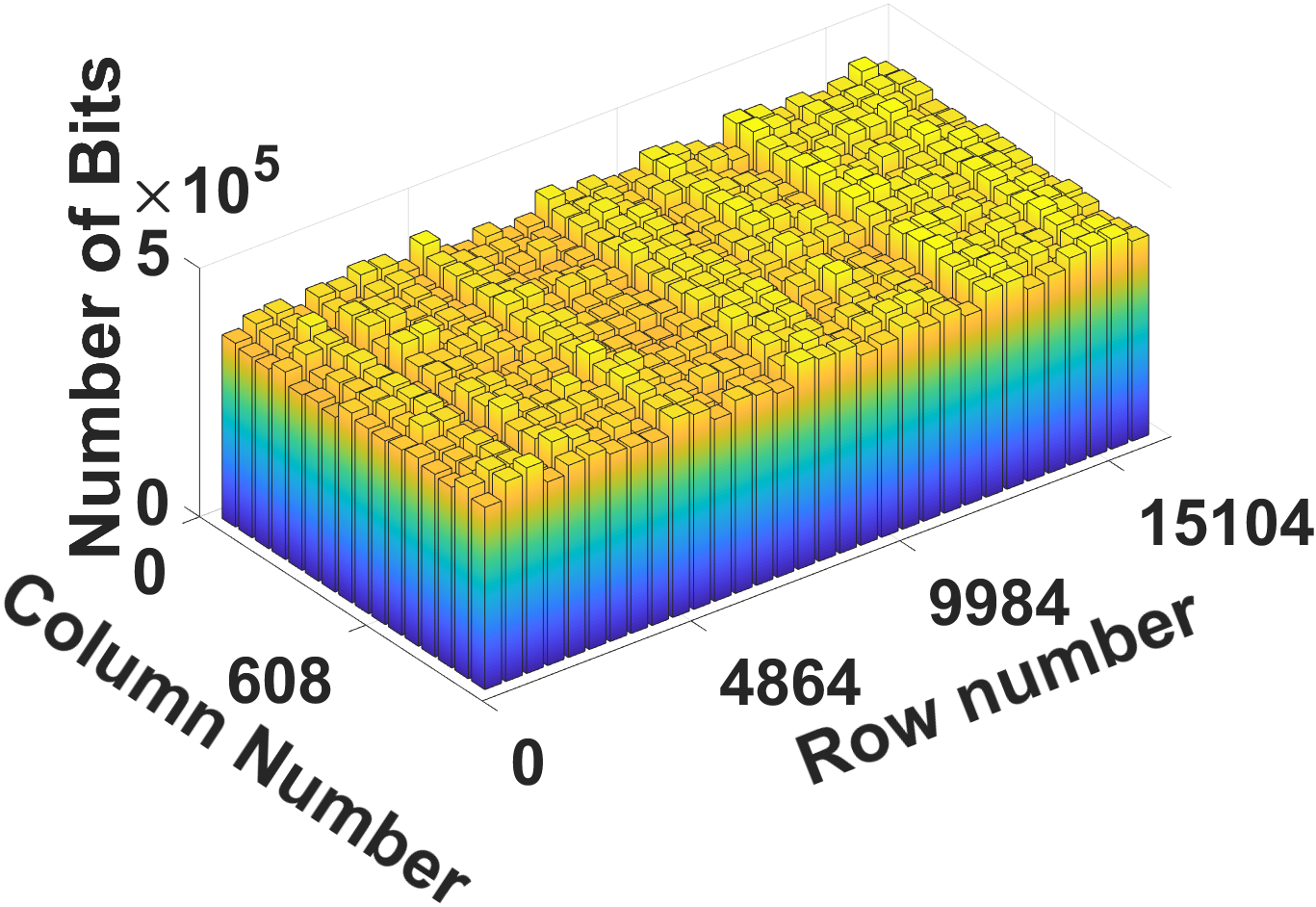} 
\caption{}
\label{fig:patIndB1}
\end{subfigure}
\captionsetup{font={sf,small},labelfont={bf,color=accessblue},justification=centering, margin=0.5cm}
\caption{Spatial location of pattern independent cells (at $t_{RP,PUF}=$ 2.5ns), (i) bit `0', (ii) bit `1'. \label{fig:patInd}}
\end{figure*}
\renewcommand\thesubfigure{\alph{subfigure}}

\begin{figure}[ht!]
\centering
\captionsetup{font={sf,small},labelfont={bf,color=accessblue},justification=centering, margin=0.5cm}
\includegraphics[width=0.48\textwidth]{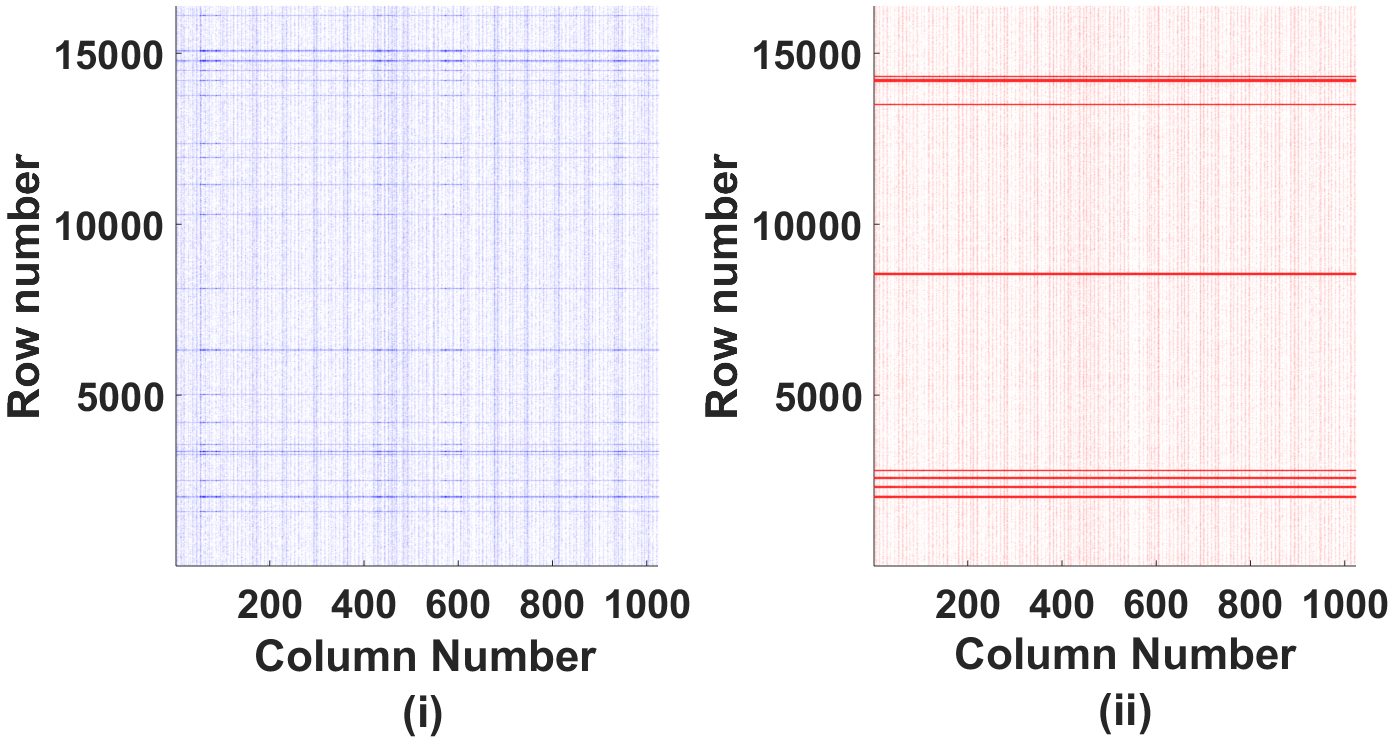} 
\caption{Pattern dependent cells (at $t_{RP,PUF}=$ 2.5ns), (i) failed to `0', and (ii) failed to `1'.}
\label{fig:patDep}
\end{figure}

\begin{figure}[ht!]
\centering
\includegraphics[width=0.48\textwidth]{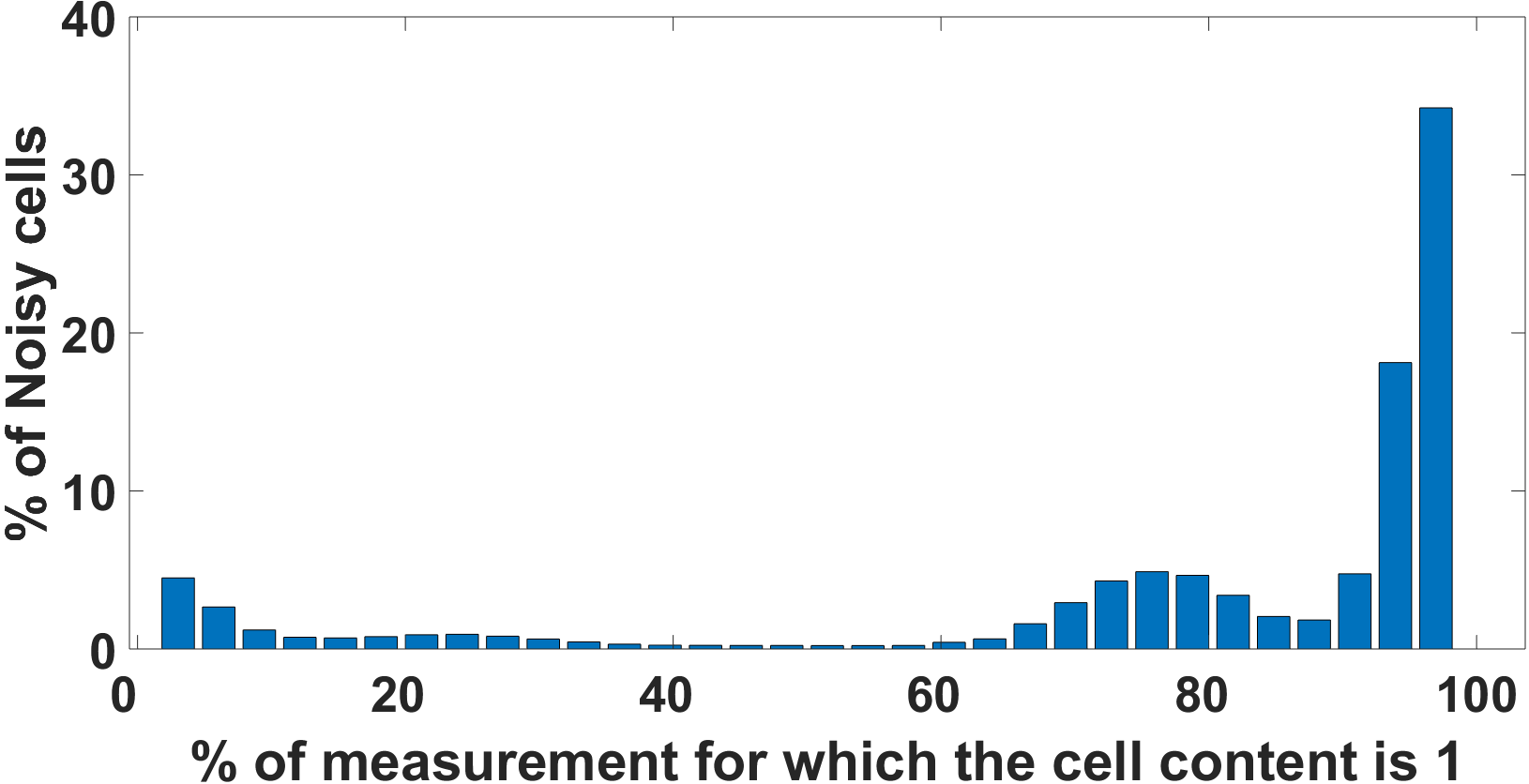}
\captionsetup{font={sf,small},labelfont={bf,color=accessblue},justification=centering, margin=0.5cm}
\caption{Noisy cell characteristics. Most of the cells are biased to `1'.}
\label{fig:noisyCell}
\end{figure}

\begin{enumerate}
\item \textbf{Pattern Independent:} Memory cells from this category always flip to a fixed value (either to `0' or to `1') regardless of the input pattern (i.e., originally written value to the DRAM cells). Fig. \ref{fig:patInd}, a 3D histogram plot to describe the spatial locality of the pattern independent cells, shows the spatial locality (along 16384 rows and 1024 columns) of output `0' (left) and `1' (right) across a random DRAM bank from vendor \textit{A}. The results show that pattern independent 1's and 0's are uniformly distributed. All memory banks from vendor \textit{B} also show similar spatial distribution (not shown in the figure). Therefore, the reduction of $t_{RP}$ is a better candidate to generate device signatures.

\item \textbf{Pattern Dependent:} The outputs of these type of cells depend on the input patterns written into the DRAM cells. The outputs are affected by the cumulative voltage of partially precharged \textit{bitline}, stored values, the coupling effect of neighbor cells, etc. We consider a memory cell as a pattern dependent if it provides different outputs for different inputs. These cells are also measurement invariant for at least one input pattern. Fig. \ref{fig:patDep} shows the DRAM cells that are dependent on input patterns \textit{0xAA}. Pattern dependent cells can be used for PUF with an enhanced challenge-response pair (CRP) space. Besides, spatial locality along both row and column are visible in Fig. \ref{fig:patDep}. The darker line in the Fig. \ref{fig:patDep} (both horizontal and vertical) represents rows and columns with the pattern dependent cells. A darker line signifies that it has more pattern dependent cells. The spatial locality might reveal the physical to logical address mapping \cite{parbor}. Fig. \ref{fig:patDep} is captured from a random bank of vendor \textit{B}, a similar type of spatial locality was found in all memory banks from all vendors. The third column (from right) in Table \ref{cellDistPre} shows the percentage of pattern dependent cells from each bank.

\item \textbf{Noisy Cells:} With partially \textit{precharged} bitlines, outputs of these cells vary from measurement to measurement. Hence, these noisy cells are not suitable to be used as PUF. The second column (from right) in Table \ref{cellDistPre} represents the percentage of noisy cells from each bank. In Fig. \ref{fig:noisyCell}, we demonstrate the distribution of noisy memory cells for a random bank from the vendor \textit{B}. The results show that the noisy cells are not entirely random (in this case, most of the cells are biased to `1'). Similar characteristics were found in other memory banks from both vendors (i.e. most of the noisy cells are biased to either `0' or `1'). Large ECC might be required if these cells are used as a PUF \cite{CIBS-ECC,ECCPUF} because of their poor reproducibility. We also found that the spatial locality of noisy cells from one bank to another is random. Therefore, a proper subset of such cells can also be used to generate a random number \cite{btalukeder:DRNG}. 
\end{enumerate}

\begin{figure}[ht!]
\centering
\includegraphics[width=0.48\textwidth]{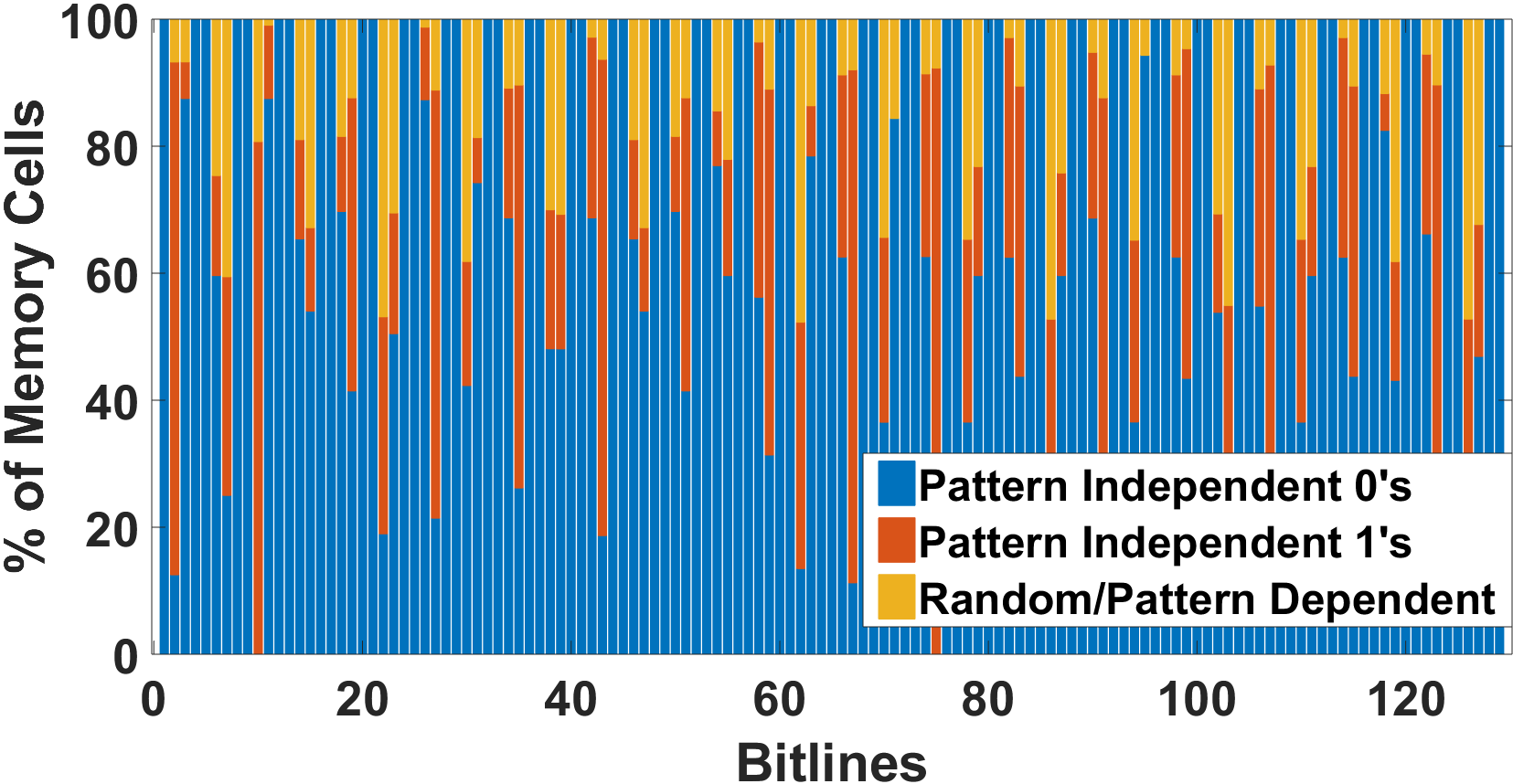}
\captionsetup{font={sf,small},labelfont={bf,color=accessblue},justification=centering, margin=0.5cm}
\caption{Cell distribution among \textit{bitlines}.}
\label{fig:pattDist}
\end{figure}
	
The complete distribution of these three types of DRAM cells along the \textit{bitline} is presented in Fig. \ref{fig:pattDist} for a given bank of \textit{vendor A}. In this figure, we presented only 128 \textit{bitlines} of two consecutive 64-bit words, where each \textit{bitline} consists of 16384 memory cells (since the total number of rows is 16384). The figure shows that all memory cells from $4n^{th}$ and $(4n+1)^{th}$ (where, $n = 1, 2, 3, ...$) bits of a word generate `0' regardless of the input patterns. One of the possible reasons is that, for these \textit{bitlines}, a large voltage difference with corresponding $\overline{bitline}s$ causes the \textit{sense-amplifier} to deviate towards a specific logic level (either `0' or `1') (see Sec. \ref{sec:divSig}).
Therefore, the generation of key from such memory cells reduces the overall entropy of the key. The proposed Algorithm \ref{selectionAlgo} eliminates such memory cells and improves the entropy. 

Table \ref{cellDistPre} summarizes the distribution of the cells of two different vendors (vendor \textit{A}, and vendor \textit{B}) at \textit{$t_{RP,PUF}$}.
The results show that more than 90\% cells from each bank of vendor \textit{A} (except the bank \textit{d}) are pattern independent while it is $<75\%$ for the vendor \textit{B}. However, we found an exception for the bank \textit{d} of vendor \textit{A} because the previously activated row fails to close at $t_{RP,2.5}$. To avoid this issue, we characterized memory cells of this bank with $t_{RP,5.0}$. For this particular memory bank, we found that the independent pattern cells are fewer in numbers than the other memory banks. We also found that the number of noisy cells increases by a significant margin than the other memory banks.

\begin{table}[ht!]
\captionsetup{singlelinecheck=false,font={sf,small},labelfont={bf,color=accessblue}}
\caption{Distribution of memory cells at the partial \textit{precharge} state.}
\centering
\setlength\tabcolsep{2.5pt}
\begin{tabular}{|c|c|c|c|c|c|c|c|}
\hline
\rule{0pt}{2ex}\multirow{3}{*}{Vendor} & \multirow{3}{*}{\begin{tabular}[c]{@{}c@{}}Memory\\ Bank ID\end{tabular}} & \multirow{3}{*}{\begin{tabular}[c]{@{}c@{}}$t_{RP,PUF}$\\ (ns)\end{tabular}} & \multicolumn{2}{c|}{\begin{tabular}[c]{@{}c@{}}Pattern\\ Independent\end{tabular}} & \multirow{3}{*}{\begin{tabular}[c]{@{}c@{}}Pattern \\ Dependent\\ (\%)\end{tabular}} & \multirow{3}{*}{\begin{tabular}[c]{@{}c@{}}Noisy\\ (\%)\end{tabular}} & \multirow{3}{*}{\begin{tabular}[c]{@{}c@{}}Vaid\\ bits\\ (\%)\end{tabular}} \\ \cline{4-5}
\rule{0pt}{2ex} &  &  & \multirow{2}{*}{0 (\%)} & \multirow{2}{*}{1 (\%)} &  &  &  \\
\rule{0pt}{2ex} &  &  &  &  &  &  &  \\ \hline
\rule{0pt}{2ex} \multirow{4}{*}{A} & a & 2.5 & 85.825 & 12.631 & 0.006 & 1.537 & 0.000 \\  \cline{2-8} 
\rule{0pt}{2ex} & b & 2.5 & 72.663 & 18.790 & 0.135 & 8.413 & 0.000 \\ \cline{2-8} 
\rule{0pt}{2ex} & c & 2.5 & 72.793 & 17.202 & 0.133 & 9.872 & 0.000 \\ \cline{2-8} 
\rule{0pt}{2ex} & d & 5 & 7.820 & 10.560 & 0.310 & 81.030 & 0.290 \\ \hline
\rule{0pt}{2ex} \multirow{2}{*}{B} & a & 2.5 & 8.226 & 63.674 & 0.519 & 27.580 & 0.001 \\ \cline{2-8} 
\rule{0pt}{2ex} & b & 2.5 & 6.339 & 53.530 & 0.113 & 40.017 & 0.001 \\ \hline
\end{tabular}
\label{cellDistPre}
\end{table}

\subsection{\textbf{PreLatPUF Evaluation:}}
We use \textit{diffuseness, uniqueness, and reliability}, three major PUF performance metrics \cite{ARO-PUF:Tauhid,ARO-PUFJournal, Maiti:metric}, to quantify and compare (with other DPUFs) the quality of the proposed PreLatPUF. The proposed Algorithm \ref{selectionAlgo}, presented in Section \ref{sec:designoverview}, is used to obtain the logical locations of the qualified memory cells. In this algorithm, we used \textit{$H_{min}$ = 0.25} and \textit{$H_{max}$ = 0.75} as the input parameters. Ideally, the Hamming distance should be 0.5. A Hamming distance of 0 represents that the PUF is not unique. We completed the registration (i.e., creating the golden data set) using the proposed Algorithm \ref{goldenDataAlgo}. We generated at least one 1024-bit key from each qualified row (or page). However, it is possible to generate multiple keys from each row since the number of qualified memory cells was more than 1024. To keep it simple, we obtained only one key from each row to test the PUF performance. The key generated from the golden data set is used as the \textit{reference key}. We refer the corresponding address for generating a \textit{reference key} as the \textit{key address}. To evaluate the performance of our proposed PreLatPUF, we created four sets of test data in four different operating conditions (will be discussed in \ref{sec:Reliability}). We measured the output for the four different input patterns (\textit{0xFF}, \textit{0xAA}, \textit{0x55}, and \textit{0x00}) and took the average.
The outputs from different operating conditions were compared with the \textit{reference key} to ensure the robustness of our proposed key generation methodology. We present the major performance metrics below. 

\subsubsection{\textbf{Diffuseness:}}
PUF device should be able to generate distinguishable responses with different challenges. For PreLatPUF, we consider the address as the challenge and corresponding cell content at reduced \textit{$t_{RP,PUF}$} as the response. To check the diffuseness, we measured inter Hamming Distance (inter HD) among the reference keys from each bank (i.e., intra-bank but inter-reference key). A 50\% of inter HD signifies that a unique key can be generated from each row (i.e., address). The average Hamming weight of 50\% also represents that the keys are random. Table \ref{tab:cellDifl} shows the average Hamming weight of each key and average Hamming distance among the different keys generated from each bank.
Though the average HD and Hamming weight for a few banks deviate from 50\%, the silicon results from all rows of each bank show that the keys generated from a distant row of the same memory bank are not repetitive.

\begin{table}[ht!]
\captionsetup{singlelinecheck=false,font={sf,small},labelfont={bf,color=accessblue}}
\caption{Average Hamming weight and average Hamming distance among the keys generated from each bank.}
\centering
\begin{tabular}{|c|c|c|c|c|}
\hline
\rule{0pt}{2ex} \multirow{3}{*}{Vendor} & \multirow{3}{*}{\begin{tabular}[c]{@{}c@{}}Memory\\ Bank ID\end{tabular}} & \multirow{3}{*}{\begin{tabular}[c]{@{}c@{}}\#Qualified\\ row (\%)\end{tabular}} & \multirow{3}{*}{\begin{tabular}[c]{@{}c@{}}Average\\ Hamming\\ Distance (\%)\end{tabular}} & \multirow{3}{*}{\begin{tabular}[c]{@{}c@{}}Average\\ Hamming\\ weight (\%)\end{tabular}} \\
\rule{0pt}{2ex} &  &  &  &  \\
\rule{0pt}{2ex} &  &  &  &  \\ \hline
\rule{0pt}{2ex} \multirow{4}{*}{A} & a & 100.00 & 48.87 & 54.23 \\ \cline{2-5} 
\rule{0pt}{2ex} & b & 92.31 & 49.35 & 53.29 \\ \cline{2-5} 
\rule{0pt}{2ex} & c & 92.30 & 49.24 & 49.24 \\ \cline{2-5} 
\rule{0pt}{2ex} & d & 67.82 & 28.97 & 53.98 \\ \hline
\rule{0pt}{2ex} \multirow{2}{*}{B} & a & 74.84 & 42.28 & 68.19 \\ \cline{2-5} 
\rule{0pt}{2ex} & b & 63.99 & 38.06 & 70.31 \\ \hline
\end{tabular}
\label{tab:cellDifl}
\end{table}

\subsubsection{\textbf{Uniqueness:}} \label{sec:Uniqueness}
Responses from different devices should be unique. This metric tells that the PUF 1 is different from the PUF 2. To quantify the uniqueness, we measured the inter Hamming Distance (inter HD) of the key from different memory banks, i.e. the HD between the two keys of two banks generated from each key address. We checked the inter HD for the following combinations by taking account following scenarios:
\begin{itemize}
\item A different pair of banks that are from the same module.
\item A different pair of banks that are from different modules but from the same vendor.
\item A different pair of banks that are from two different vendors.
\end{itemize}

\begin{figure}[ht!]
\centering
\captionsetup{justification=centering, margin= .6cm}
\begin{subfigure}[t!]{0.22\textwidth}
\includegraphics[width=\textwidth]{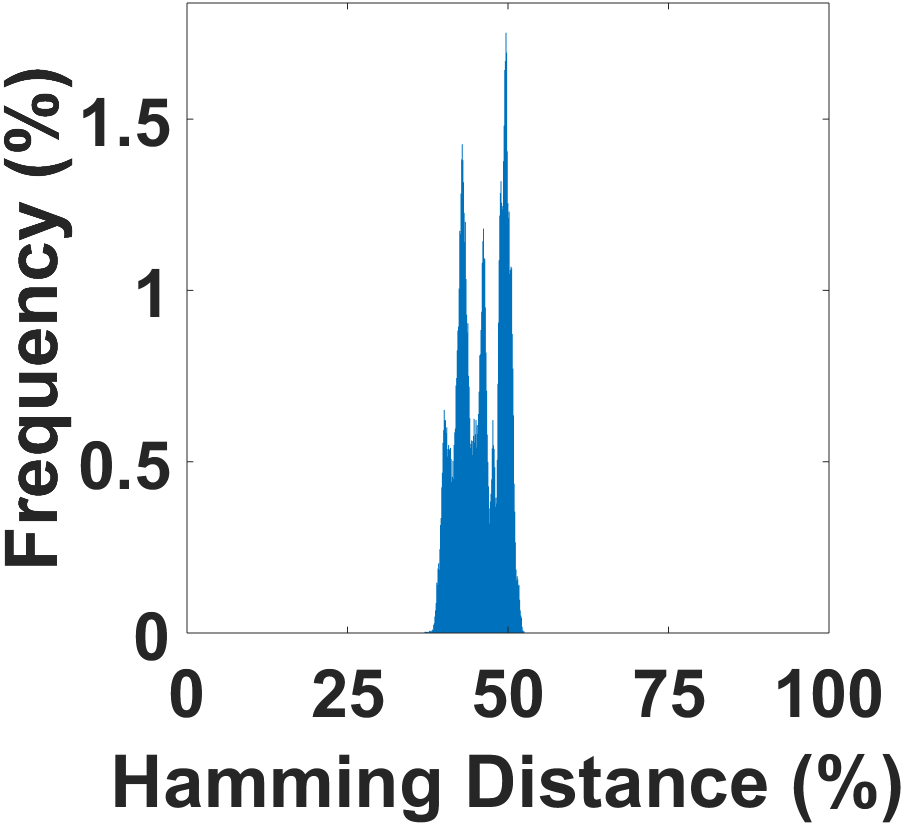} 
\caption{}
\label{fig:interHD_mb4}
\end{subfigure}
~
\begin{subfigure}[t!]{0.22\textwidth}
\includegraphics[width=\textwidth]{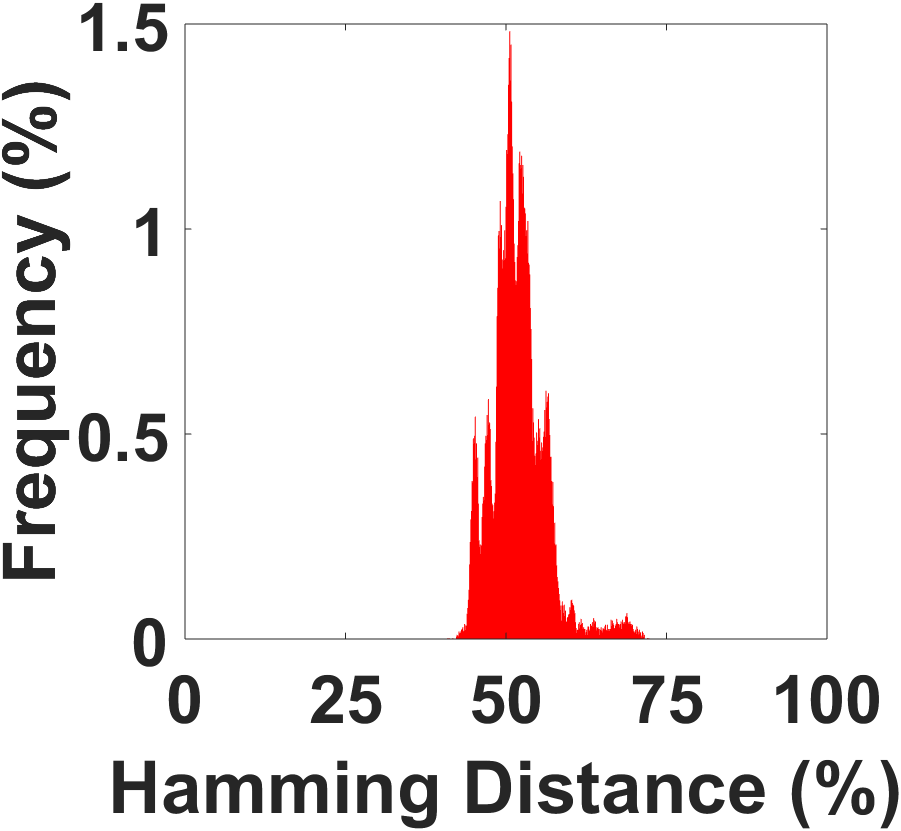}
\caption{}
\label{fig:interHD_sb2}
\end{subfigure}
\captionsetup{font={sf,small},labelfont={bf,color=accessblue},justification=centering, margin=0cm}
\caption{Inter Hamming distance (at $t_{RP,PUF}=$ 2.5ns) for the worst case from (i) vendor \textit{A} and (ii) vendor \textit{B}. \label{interHD}}
\end{figure}

Fig. \ref{interHD} shows the inter HD at the worst case (i.e., the largest deviation from 50\% inter HD) scenario for both vendors. In the worst scenario, for the vendor \textit{A}, the average, minimum, and maximum inter HD are 45.78\%, 37.05\%, and 52.5\% respectively. For the vendor \textit{B}, the mean, minimum, and maximum inter HD are 51.91\%, 40.92\%, and 72.23\%, respectively. Therefore, we can conclude that the key generated from the proposed PreLatPUF is unique.

\renewcommand\thesubfigure{\roman{subfigure}}
\renewcommand\thesubfigure{\alph{subfigure}}

\subsubsection{\textbf{Reliability:}} \label{sec:Reliability}
\begin{figure*}[ht!]
\centering
\begin{subfigure}[t!]{0.49\textwidth}
\includegraphics[width=\textwidth]{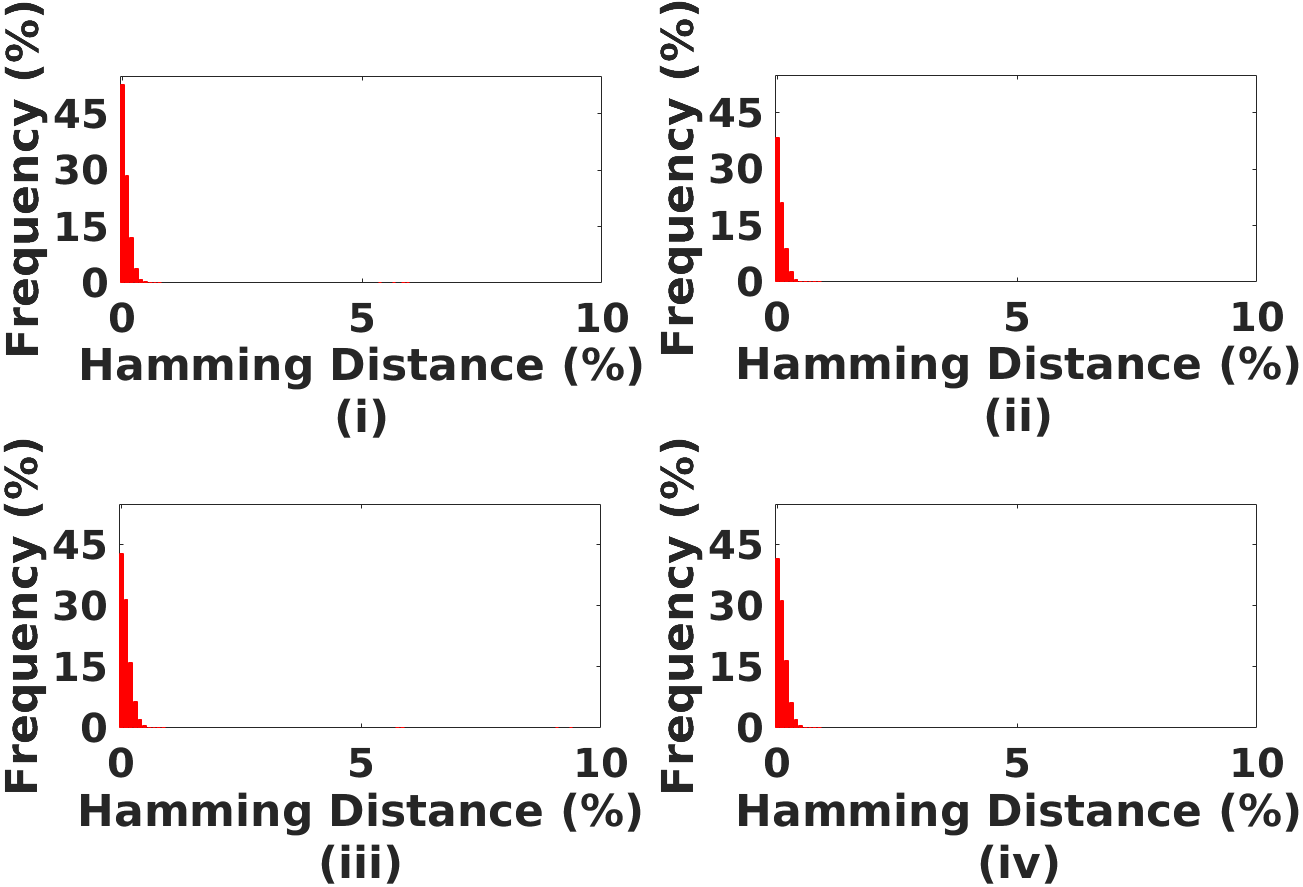} 
\caption{}
\label{fig:intraHD_mb4}
\end{subfigure}
~
\begin{subfigure}[t!]{0.49\textwidth}
\includegraphics[width=\textwidth]{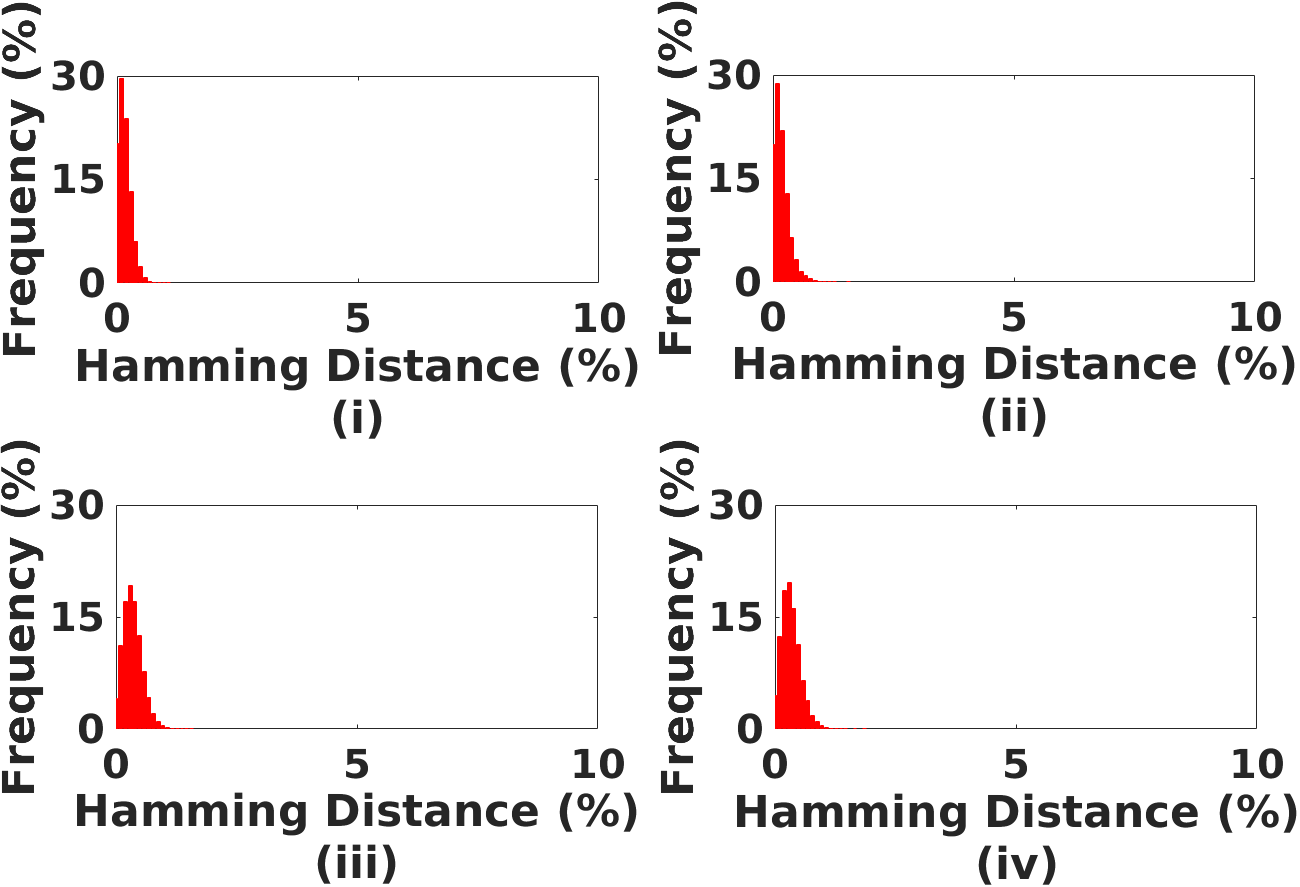}
\caption{}
\label{fig:intraHD_sb2}
\end{subfigure}
\captionsetup{font={sf,small},labelfont={bf,color=accessblue},justification=centering, margin=1.2cm}
\caption{Intra HD for the worst case from- (a) vendor \textit{A} (at $t_{RP,PUF}=$ 2.5ns), (b) vendor \textit{B} (at $t_{RP,PUF}=$ 2.5ns) with (i) NVRT, (ii) LVRT, (iii) HVRT, and (iv) NVHT. \label{intraHD}}
\end{figure*}

Same response (i.e., PUF output) should be generated to its entire lifetime at any operating condition. The reproducibility at different operating conditions is presented in Fig. \ref{intraHD}. This figure presents only the worst results from each vendor (i.e., memory bank with the most significant deviation from 0\% intra HD). To examine the robustness of the proposed PreLatPUF at extreme operating conditions, we collected results at four different operating conditions: (i) nominal voltage and room temperature (NVRT), (ii) low-voltage and room temperature (LVRT), (iii) high-voltage and room temperature (HVRT), and (iv) nominal voltage and high temperature (NVHT). The results show that the memory module from vendor \textit{A} is less robust than the vendor \textit{B} at the reduced operating voltage. For vendor \textit{A}, we can only change the operating voltage by \textit{$-$20mv} without causing an excessive error on PUF response. On the other hand, the DRAM module from vendor \textit{B} can tolerate \textit{$-$55mv} change in operating voltage. Table \ref{tab:intraHD} presents the intra HD at different operating conditions. The column 4 of Table \ref{tab:intraHD} represents change in operating voltage from the nominal (1.5V) and the column 5 represents change in temperature from room temperature (25$\degree$C). The results show that all memory banks from both vendors are robust against temperature variations.

The rightmost column of Table \ref{tab:intraHD} shows that the robustness of the PUF output improves as we increase the operating voltage for those banks that possess dominant pattern independent `0' cells at the reduced $t_{RP}$. An increase in the voltage makes these cells more immune to noise. On the other hand, the banks with dominant pattern independent `1' cells show the opposite behavior (i.e., the robustness of the PUF output increases as we reduce the operating voltage). A decrease in the voltage makes these cells more immune to noise. However, the bank \textit{d} from the vendor \textit{A} produces a slight robust output with the change in the voltage (increased or decreased) because this bank produces noisier cells than other banks. 

\begin{table}[ht!]
\captionsetup{singlelinecheck=false,font={sf,small},labelfont={bf,color=accessblue}}
\caption{Intra HD at different operating conditions.}
\centering
\setlength\tabcolsep{2.5pt}
\begin{tabular}{|c|c|c|c|c|c|c|c|c|}
\hline
\rule{0pt}{2ex} \multirow{3}{*}{Vendor} & \multirow{3}{*}{\begin{tabular}[c]{@{}c@{}}Memory\\ Bank ID\end{tabular}} & \multirow{3}{*}{\begin{tabular}[c]{@{}c@{}}Operating\\ Codition\end{tabular}} & \multirow{3}{*}{\begin{tabular}[c]{@{}c@{}}$\Delta{V}$\\ (mV)\end{tabular}} & \multirow{3}{*}{\begin{tabular}[c]{@{}c@{}}$\Delta{T}$\\ ($\degree$ C)\end{tabular}} & \multicolumn{2}{c|}{\multirow{2}{*}{Intra HD}} & \multicolumn{2}{c|}{\multirow{2}{*}{\begin{tabular}[c]{@{}c@{}}Key with\\ Intra HD\end{tabular}}} \\
\rule{0pt}{2ex} &  &  &  &  & \multicolumn{2}{c|}{} & \multicolumn{2}{c|}{} \\ \cline{6-9} 
\rule{0pt}{2ex} &  &  &  &  & $\mu$ & $\sigma$ & $>$ 1\% & $>$ 30\% \\ \hline
\rule{0pt}{2ex} \multirow{16}{*}{A} & \multirow{4}{*}{a} & NVRT & 0 & 0 & 0.48 & 0.07 & 0.00 & 0.00 \\ \cline{3-9} 
\rule{0pt}{2ex} &  & LVRT & $-$20 & 0 & 0.05 & 0.08 & 0.00 & 0.00 \\ \cline{3-9} 
\rule{0pt}{2ex} &  & HVRT & $+$55 & 0 & 0.07 & 0.09 & 0.00 & 0.00 \\ \cline{3-9} 
\rule{0pt}{2ex} &  & NVHT & 0 & $+$20 & 0.06 & 0.09 & 0.00 & 0.00 \\ \cline{2-9} 
\rule{0pt}{2ex} & \multirow{4}{*}{b} & NVRT & 0 & 0 & 0.47 & 3.17 & 1.57 & 0.00 \\ \cline{3-9} 
\rule{0pt}{2ex} &  & LVRT & $-$20 & 0 & 2.94 & 10.55 & 7.81 & 2.91 \\ \cline{3-9} 
\rule{0pt}{2ex} &  & HVRT & $+$55 & 0 & 0.09 & 0.10 & 0.00 & 0.00 \\ \cline{3-9} 
\rule{0pt}{2ex} &  & NVHT & 0 & $+$20 & 0.67 & 3.84 & 2.34 & 0.01 \\ \cline{2-9} 
\rule{0pt}{2ex} & \multirow{4}{*}{c} & NVRT & 0 & 0 & 0.49 & 3.34 & 1.54 & 0.03 \\ \cline{3-9} 
\rule{0pt}{2ex} &  & LVRT & $-$20 & 0 & 7.77 & 12.38 & 27.95 & 0.46 \\ \cline{3-9} 
\rule{0pt}{2ex} &  & HVRT & $+$55 & 0 & 0.09 & 0.12 & 0.01 & 0.00 \\ \cline{3-9} 
\rule{0pt}{2ex} &  & NVHT & 0 & $+$20 & 0.52 & 3.38 & 1.54 & 0.02 \\ \cline{2-9} 
\rule{0pt}{2ex} & \multirow{4}{*}{d} & NVRT & 0 & 0 & 1.54 & 9.02 & 4.37 & 2.74 \\ \cline{3-9} 
\rule{0pt}{2ex} &  & LVRT & $-$20 & 0 & 1.69 & 8.87 & 8.87 & 2.66 \\ \cline{3-9} 
\rule{0pt}{2ex} &  & HVRT & $+$55 & 0 & 1.47 & 8.73 & 4.29 & 2.64 \\ \cline{3-9} 
\rule{0pt}{2ex} &  & NVHT & 0 & $+$20 & 4.72 & 8.36 & 9.35 & 2.62 \\ \hline
\rule{0pt}{2ex} \multirow{8}{*}{B} & \multirow{4}{*}{a} & NVRT & 0 & 0 & 1.97 & 10.25 & 3.37 & 3.25 \\ \cline{3-9} 
\rule{0pt}{2ex} &  & LVRT & $-$55 & 0 & 2.11 & 10.19 & 3.36 & 3.17 \\ \cline{3-9} 
\rule{0pt}{2ex} &  & HVRT & $+$55 & 0 & 1.92 & 10.02 & 3.53 & 3.17 \\ \cline{3-9} 
\rule{0pt}{2ex} &  & NVHT & 0 & $+$20 & 2.13 & 10.23 & 3.76 & 3.26 \\ \cline{2-9} 
\rule{0pt}{2ex} & \multirow{4}{*}{b} & NVRT & 0 & 0 & 1.93 & 10.55 & 3.24 & 2.62 \\ \cline{3-9} 
\rule{0pt}{2ex} &  & LVRT & $-$55 & 0 & 2.22 & 10.30 & 5.68 & 2.52 \\ \cline{3-9} 
\rule{0pt}{2ex} &  & HVRT & $+$55 & 0 & 1.95 & 10.35 & 3.18 & 2.53 \\ \cline{3-9} 
\rule{0pt}{2ex} &  & NVHT & 0 & $+$20 & 1.99 & 10.55 & 3.39 & 2.74 \\ \hline
\end{tabular}
\label{tab:intraHD}
\end{table}

\subsection{\textbf{Performance Comparison}}
\subsubsection{\textbf{Evaluation Time:}}

We use two approaches to quantify and compare (with other DPUFs) the evaluation time. Eq. \eqref{eq3} (the first approach) and Eq. \eqref{eq4} (the second approach) are used to compare the time overhead required for the Key generation. The first approach measures the time required to receive the response after sending the challenge from the host. The second approach, on the other hand, is intended to measure the required time to produce the key in the evaluation board ($t_{exec}$)\footnote{The current implementation does not support a separate measurement of $t_{exec}$ and $t_{host\_receive}$.}. 

\begin{flalign}\label{eq3}
&\ \ \ \ \mathcal{T}_{eval1} \approx t_{host\_send} + t_{exec} + t_{host\_receive} + t_{store}&
\end{flalign}
\begin{flalign}\label{eq4}
&\ \ \ \ \mathcal{T}_{eval2} \approx t_{exec} + t_{host\_receive} &
\end{flalign}
where,\\
$\mathcal{T}_{eval1} =$ evaluation with the first approach,\\
$\mathcal{T}_{eval2} =$ evaluation with the second approach,\\
$t_{host\_send} =$ time required to send the command to the evaluation board from the host computer,\\
$t_{exec} = $ time required to execute the command in the evaluation board,\\
$t_{host\_receive} =$ time required to send back the read data to the host computer from the evaluation board, and\\
$t_{store} = $ time required to store the read data to a storage device.\\

With the first approach, the worst average time is 1.59ms (worst among all banks, see Table \ref{evalTime}), which is 74us with the second approach. However, the evaluation time can be measured more accurately by inserting a local counter inside the FPGA. Note that we did not include the characterization phase in the evaluation time (see Eq. \eqref{eq3} and Eq. \eqref{eq4}) since the cell characterization is performed once during the registration. We, also, did not include the time to write a specific data pattern because we did not consider pattern dependent cells in this paper. However, the writing time needs to be added if pattern dependent cells are considered to generate a large CRP space. The average system-level evaluation time of reduced \textit{$t_{RCD}$}-based DPUF is 88.2ms \cite{DRAMLatencyPUF}, which is still $\sim1,192X$ slower (considering the worst evaluation time with the second approach) than our proposed method. On the other hand, the retention-based DPUF takes order of minutes to generate a device signature with enough retention failures \cite{retentionPUF}. 

\begin{table}[ht!]
\captionsetup{singlelinecheck=false,font={sf,small},labelfont={bf,color=accessblue}}
\caption{Average PreLatPUF evaluation time.}
\centering
\setlength\tabcolsep{10pt} 
\begin{tabular}{|c|c|c|c|}
\hline
\rule{0pt}{2ex} \multirow{2}{*}{Vendor} & \multirow{2}{*}{\begin{tabular}[c]{@{}c@{}}Memory\\ Bank ID\end{tabular}} & \multirow{2}{*}{\begin{tabular}[c]{@{}c@{}}\#Required\\ Burst (mean)\end{tabular}} & \multirow{2}{*}{\begin{tabular}[c]{@{}c@{}}Mean\\ Evaluation time (ms)\end{tabular}} \\
 &  &  &  \\ \hline
\rule{0pt}{2ex} \multirow{4}{*}{A} & a & 9.00 & 0.51 \\ \cline{2-4} 
\rule{0pt}{2ex} & b & 6.43 & 0.41 \\ \cline{2-4} 
\rule{0pt}{2ex} & c & 7.19 & 0.47 \\ \cline{2-4} 
\rule{0pt}{2ex} & d & 16.10 & 0.93 \\ \hline
\rule{0pt}{2ex} \multirow{2}{*}{B} & a & 28.15 & 1.59 \\ \cline{2-4} 
 & b & 24.18 & 1.34 \\ \hline
\end{tabular}
\label{evalTime}
\end{table}

\subsubsection{\textbf{System Level Disruption:}}
For most of the DRAM modules, the granularity of refreshing the DRAM contents is rank. Therefore, we need to increase the refresh interval for entire memory rank for evaluating retention-based DPUF. As a result, it causes random data corruption over the whole rank. Also, due to the long evaluation time of the retention-based DPUF, the particular DRAM rank becomes unavailable for other applications for a long time. In the proposed PreLatPUF, the reduced \textit{$t_{RP}$} only affects the cells that are being accessed. We also checked the interference to the neighborhood rows of the target row that is being accessed for key generation. To do so, we arbitrarily selected consecutive 1000 rows from each memory bank. Then, we read the data from all odd-numbered rows at the \textit{$t_{RP,PUF}$} and investigated the impact on the memory cells of the even-numbered row with nominal \textit{$t_{RP}$}. Our results show that there is no data corruption in the adjacent rows.

However, though the latency-based DPUF of \cite{DRAMLatencyPUF} at the reduced $t_{RCD}$ is fast, this type of DPUF evaluation needs a filtering mechanism upon each access, which causes both computational and hardware overheads. In our proposed mechanism, we register the eligible PUF cells once in its entire lifetime (see Section \ref{sec:cellSelection}). Once the suitable cells for PUF operation are determined, the evaluation of our proposed PUF is straight-forward (i.e. request the response by sending an address and then compare only the eligible cells' content with the golden data). The proposed PUF evaluation has the least evaluation time. Therefore, the proposed PreLatPUF can be used in run-time, which is impossible in many existing DPUFs \cite{retentionPUF, retentionPUF2, Keller:DRAMPUF}.

\subsubsection{\textbf{Robustness:}}
The robustness (i.e., the effect of different operating conditions and environmental variations) of the proposed PreLatPUF is shown in Table \ref{tab:intraHD}. The impact of operating voltage and temperature variations in DPUFs have been explored before \cite{DRAMLatencyPUF, retentionPUF}. In this paper, we compared the robustness between the proposed PreLatPUF and retention-based DPUF at different operating conditions. To accumulate the retention-based failures, we chose a random memory segment with 1000 rows from each bank. At first, we stored logic `1' to all memory cells under the segment, and then the refresh interval was prolonged until we obtained at least $\sim$2\% failures at the NVRT. For a specific bank, same refresh interval was maintained for all other operating conditions. For the proposed PreLatPUF, we measured the data errors with four input patterns (\textit{0xFF}, \textit{0xAA}, \textit{0x55}, and \textit{0x00}) at \textit{$t_{RP,PUF}$} for the same 1000 rows. We used the \textit{Jaccard Index} to compare the robustness of our proposed PreLatPUF with the retention-based PUF. For the retention-based DPUF, the PUF characteristics are evaluated from the location of the failed bits. For example, in our case, retention-based failed bits are always failed from logic `1' to logic `0'. But the location of the failed bits differs from one device to another. For the two sets of the measurements ($M_1$, $M_2$), the \textit{Jaccard Index} is measured as $\frac{M_1 \cap M_2}{M_1 \cup M_2}$, where $M_1 \cap M_2$ is the total matched failed bits and $M_1 \cup M_2$ is the total failed bits from two measurements $M_1$ and $M_2$ \cite{DRAMLatencyPUF, Schaller:hammerPUF}. For better reproducibility, the intra \textit{Jaccard Index} should be $\sim$1. Table \ref{compJindex} shows the comparison between PreLatPUF and retention-based PUF. The results show that the proposed PreLatPUF is more robust than the retention-based DPUF. The retention-based PUF is more susceptible to the temperature variation compared to the PreLatPUF. This is because the retention-based failed bit is mostly emphasized by the charge leakage rate of DRAM cells, which has a strong exponential dependence on the temperature \cite{retentionPUF, retentionPUF2, retentionPUF3, retBehavior, Hassan:ChargeCache, leakagedistribution}. On the other hand, the change in \textit{$t_{RP}$} is very negligible as temperature changes. The \textit{$t_{RP}$} changes only ($\sim$3\%) as temperature changes from 27$\degree$C to 85$\degree$C \cite{runtimePerformanceDRAM}.

For the $t_{RCD}$-based DPUF, the results shown in \cite{DRAMLatencyPUF} suggest that it can tolerate only a small change in temperature (e.g., 5$\degree$C). On the other hand, for the PreLatPUF, we increased the temperature by 20$\degree$ and found a negligible change in the signature. The results presented in \cite{runtimePerformanceDRAM} also suggest that the temperature dependency of $t_{RCD}$ is stronger than the temperature dependency of $t_{RP}$.

\begin{table}[ht!]
\captionsetup{singlelinecheck=false,font={sf,small},labelfont={bf,color=accessblue}}
\caption{\textit{Jaccard Index} at different operating conditions for the PreLatPUF and the retention-based DPUF. }
\centering
\begin{tabular}{|c|c|c|c|c|}
\hline
\rule{0pt}{2ex} \multirow{3}{*}{Vendor} & \multirow{3}{*}{\begin{tabular}[c]{@{}c@{}}Memory\\ Bank ID\end{tabular}} & \multirow{3}{*}{$M_1$, $M_2$} & \multicolumn{2}{c|}{Jaccard Index} \\ \cline{4-5} 
\rule{0pt}{2ex} &  &  & \multirow{2}{*}{\begin{tabular}[c]{@{}c@{}}Proposed\\ PreLatPUF\end{tabular}} & \multirow{2}{*}{\begin{tabular}[c]{@{}c@{}}Retention\\ Based DPUF\end{tabular}} \\
 &  &  &  &  \\ \hline
\rule{0pt}{2ex} \multirow{12}{*}{A} & \multirow{3}{*}{a} & NVRT, LVRT & 0.997 & 0.926 \\ \cline{3-5} 
\rule{0pt}{2ex} &  & NVRT, HVRT & 0.997 & 0.968 \\ \cline{3-5} 
\rule{0pt}{2ex} &  & NVRT, NVHT & 0.997 & 0.349 \\ \cline{2-5} 
\rule{0pt}{2ex} & \multirow{3}{*}{b} & NVRT, LVRT & 0.980 & 0.902 \\ \cline{3-5} 
\rule{0pt}{2ex} &  & NVRT, HVRT & 0.997 & 0.970 \\ \cline{3-5} 
\rule{0pt}{2ex} &  & NVRT, NVHT & 0.986 & 0.356 \\ \cline{2-5} 
\rule{0pt}{2ex} & \multirow{3}{*}{c} & NVRT, LVRT & 0.929 & 0.930 \\ \cline{3-5} 
\rule{0pt}{2ex} &  & NVRT, HVRT & 0.997 & 0.960 \\ \cline{3-5} 
\rule{0pt}{2ex} &  & NVRT, NVHT & 0.985 & 0.355 \\ \cline{2-5} 
\rule{0pt}{2ex} & \multirow{3}{*}{d} & NVRT, LVRT & 0.994 & 0.941 \\ \cline{3-5} 
\rule{0pt}{2ex} &  & NVRT, HVRT & 0.983 & 0.968 \\ \cline{3-5} 
\rule{0pt}{2ex} &  & NVRT, NVHT & 0.996 & 0.279 \\ \hline
\rule{0pt}{2ex} \multirow{6}{*}{B} & \multirow{3}{*}{a} & NVRT, LVRT & 0.968 & 0.962 \\ \cline{3-5} 
\rule{0pt}{2ex} &  & NVRT, HVRT & 0.961 & 0.847 \\ \cline{3-5} 
\rule{0pt}{2ex} &  & NVRT, NVHT & 0.968 & 0.421 \\ \cline{2-5} 
\rule{0pt}{2ex} & \multirow{3}{*}{b} & NVRT, LVRT & 0.965 & 0.952 \\ \cline{3-5} 
\rule{0pt}{2ex} &  & NVRT, HVRT & 0.968 & 0.950 \\ \cline{3-5} 
\rule{0pt}{2ex} &  & NVRT, NVHT & 0.971 & 0.457 \\ \hline
\end{tabular}
\label{compJindex}
\end{table}

\section{Conclusion} \label{sec:conclusion}
In this paper, we proposed a DRAM-based PUF that exploits the precharge-latency variations in DRAM cells. We characterized DRAM cells' errors at the reduce precharge-latency to find the most suitable DRAM cells in order to produce random, unique, and reliable device signatures. The silicon results from commercially available DRAM modules show that the proposed device signature scheme and algorithm can generate robust PUF outputs at a much faster rate.

\appendices
\section{Impact of Reduced \textit{Activation} time} \label{app:A}
\begin{figure}[ht!]
\centering
\includegraphics[width=0.48\textwidth]{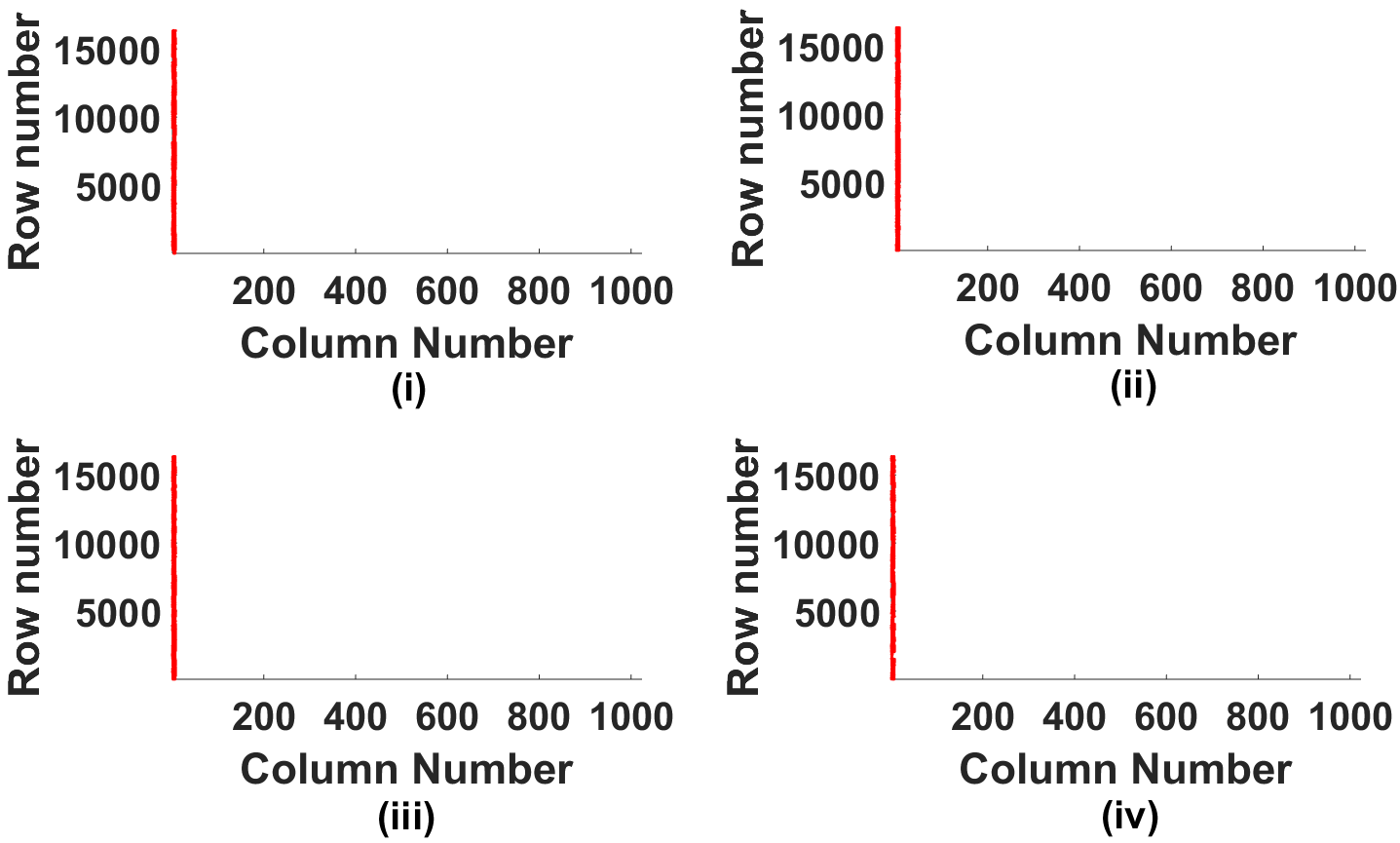}
\captionsetup{font={sf,small},labelfont={bf,color=accessblue},justification=centering, margin=0.5cm}
\caption{Failed bits at \textit{$t_{RCD,5.0}$} with input pattern- (i) \textit{0x00}, (ii) \textit{0x55}, (iii) \textit{0xAA}, and (iv) \textit{0xFF}.}
\label{fig:actTime}
\end{figure}

In figure \ref{fig:actTime}, red spots represent the failed bits at the reduced activation time (\textit{$t_{RCD,5.0}$}) for a DRAM bank. The results show that the failed bits are only observed at the first accessed cache line (i.e., just in the first column). A similar observation was concluded in \cite{DRAMLatencyPUF} and \cite{Kevin:Latency}.

\section*{Acknowledgment}
 We would like to thank Hasan Hassan (ETH Z{\"u}rich) \& CMU for the SoftMC software.


\begin{IEEEbiography}[{\includegraphics[width=1in,height=1.25in,clip,keepaspectratio]{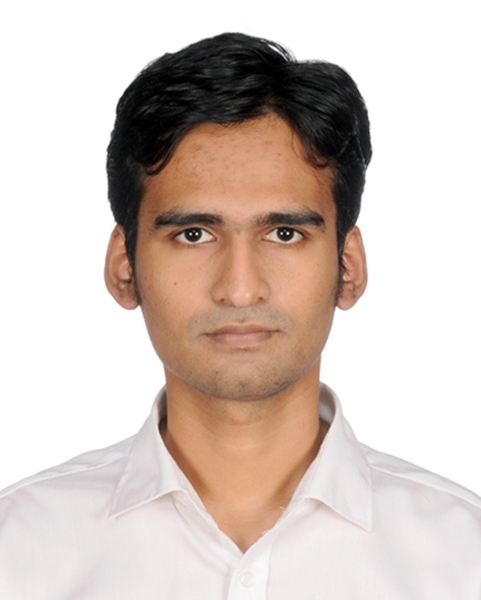}}]{B. M. S. Bahar Talukder} (S'18) is a Ph.D. student in Computer Engineering at the University of Alabama in Huntsville. He received his Bachelor's degree from Bangladesh University of Engineering and Technology, Dhaka, Bangladesh. His primary research interests include hardware security, secured computer architecture, and emerging memory technologies.
\end{IEEEbiography}

\begin{IEEEbiography}[{\includegraphics[width=1in,height=1.25in,clip,keepaspectratio]{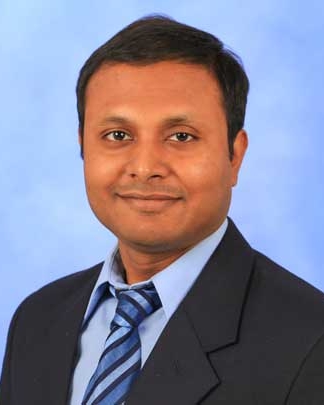}}]{Biswajit Ray} (S'12-M'16) is an Assistant Professor of Electrical and Computer Engineering with the University of Alabama in Huntsville, AL, USA, where he leads the Hardware Reliability Lab. Dr. Ray received Ph.D. from Purdue University, West Lafayette, IN in 2013 and then he worked in SanDisk Corporation, California, USA, developing 3D NAND flash memory technology. His current research spans the boundaries of electron devices and systems for addressing the challenges in hardware security and reliability. Dr. Ray has 8 U.S. issued patents on non-volatile memory devices and systems, published more than 40 research papers in international journals and conferences. 
\end{IEEEbiography}

\begin{IEEEbiography}[{\includegraphics[width=1in,height=1.25in,clip,keepaspectratio]{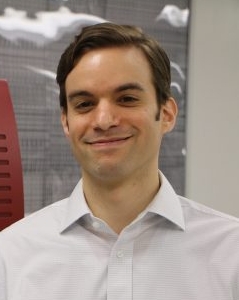}}]{Domenic Forte} (S'09-M'13-SM'18) received the B.S. degree in electrical engineering from the Manhattan College, Riverdale, NY, USA, in 2006, and the M.S. and Ph.D. degrees in electrical engineering from the University of Maryland at College Park, College Park, MD, USA, in 2010 and 2013, respectively. He is currently an Assistant Professor with the Electrical and Computer Engineering Department, University of Florida, Gainesville, FL, USA. His current research interests include the domain of hardware security, including the investigation of hardware security primitives, hardware Trojan detection and prevention, electronics supply-chain security, and antireverse engineering. Dr. Forte was a recipient of the Young Investigator Award from the Army Research Office, the NSF CAREER Award, and the George Corcoran Memorial Outstanding Teaching Award from the Electrical and Computer Engineering Department, University of Maryland. 
\end{IEEEbiography}


\begin{IEEEbiography}[{\includegraphics[width=1in,height=1.25in,clip,keepaspectratio]{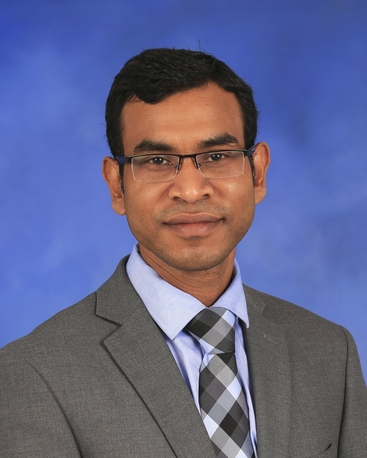}}]{Md Tauhidur Rahman} (S'12-M'18) is an assistant professor at the Electrical and Computer Engineering Department in the University of Alabama in Huntsville, USA, where he directs the SeRLoP Research Lab. His current research interests include hardware security and trust, security of memory organization and modern architectures, embedded security, and reliability. He obtained his Ph.D. degree from the University of Florida in 2017 and master's degree from the University of Connecticut in 2015. Dr. Rahman received the NSF CRII Award in 2019.
\end{IEEEbiography}

\EOD

\end{document}